\documentclass[prb, 12pt, superscriptaddress]{revtex4} 
\usepackage{amssymb}
\usepackage{amsmath}
\usepackage{bbm}
\usepackage{graphicx}
\usepackage{color}

\begin{document}

\author{Philipp Werner}
\affiliation{Department of Physics, University of Fribourg, 1700 Fribourg, Switzerland}
\author{Martin Eckstein}
\affiliation{Max Planck Research Department for Structural Dynamics, University of Hamburg-CFEL, 22761 Hamburg, Germany}

\title{Effective doublon and hole temperatures in the photo-doped dynamic Hubbard model}

\date{August 28, 2015}

\hyphenation{}

\begin{abstract} 
Hirsch's dynamic Hubbard model describes the effect of orbital expansion with occupancy by coupling the doublon operator to an auxiliary boson. We use the nonequilibrium dynamical mean field method to study the properties of doublon and hole carriers in this model in the strongly correlated regime. In particular, we discuss how photodoping leads to doublon and hole populations with different effective temperatures, and we analyze the relaxation behavior as a function of the boson coupling and boson energy. In the polaronic regime, the nontrivial energy exchange between doublons, holes and bosons can result in a negative temperature distribution for the holes. 
\end{abstract}


\maketitle

\section{Introduction}

Strongly correlated electron systems are often described in terms of the Hubbard model or its multiorbital extensions. This model assumes that the electrons interact only locally on each lattice site, and that this interaction is of a particle-hole symmetric form. In the case of the single-band Hubbard model, the interaction on site $i$, parametrized by $U$, can be written as $U(n_{i\uparrow}-\tfrac12)(n_{i\downarrow}-\tfrac12)$, so that the energy costs for adding and removing an electron in the half-filled state are the same. 
As was pointed out by Hirsch and co-workes in a series of papers, this model ignores the fact that doubly occupied atomic orbitals expand as a result of Coulomb interactions and electron correlations. To capture this physics, Hirsch introduced the dynamic Hubbard model,\cite{Hirsch2001} which contains an additional coupling of the doublon operator to an auxiliary boson. The expansion of the orbital and associated reduction of the Coulomb interaction is then described by the dynamics of the boson and the resulting screening effect. Several variants of this model have subsequently been formulated and explored. \cite{Hirsch2002a, Hirsch2002b, Hirsch2002c, Hirsch2003} These equilibrium studies have revealed interesting properties resulting from the filling-dependent particle-hole asymmetric correlation effects, which may even be relevant for the understanding of unconventional superconductivity.\cite{Hirsch2002b, Marsiglio2003}  

More recently, a dynamic Hubbard model has been proposed to explain the behavior of organic charge-transfer salts in which molecular orbitals are periodically modulated via the selective excitation of an optical phonon mode with intense laser fields. \cite{Kaiser2014, Singla2014} In this context, the particle-hole asymmetric electron-boson coupling provides an experimental ``knob" to tune the interaction effects in the solid, and hence raises the interesting prospect of dynamical control of material properties.  

Theoretically, the nonequilibrium properties of dynamic Hubbard models are largely unexplored, but as in equilibrium, one can expect interesting effects resulting from the particle-hole asymmetric correlations. 
Examples are the nontrivial interplay between heat and particle currents (Seebeck effect), or the unusual nature of the photo-doped insulator with its coexisting weakly/strongly correlated doublons/holes. Motivated by these considerations, we investigate here some nonequilibrium phenomena in the  dynamic Hubbard model by means of the nonequilibrium dynamical mean field (DMFT) formalism.\cite{Aoki2014} 
In particular, we will study how the dynamic Coulomb correlations are reflected in the relaxation and recombination of doublons and holes produced by 
an impulsive stimulation of the Mott insulating state. We will show that the particle-hole asymmetry leads to different transient temperatures for the doublons and holes during the relaxation, and that the respective cooling rates depend in a nontrivial way on the coupling strength of the auxiliary boson. 

The rest of this paper is organized as follows. Section~\ref{model} describes the model considered in this study and the approximate impurity solver used in the DMFT calculations. Section~\ref{results} presents results for the nonequilibrium spectral function and the time-evolution of the effective doublon and hole temperatures, while Sec.~\ref{summary} provides a short summary.

\section{Model and Method}
\label{model}

We consider the dynamic Hubbard model introduced in the original paper by Hirsch,\cite{Hirsch2001}
\begin{equation}
H=\sum_{i,j,\sigma} v_{ij} c^\dagger_{i\sigma}c_{j\sigma}+\sum_i \Big[ U n_{i\uparrow} n_{i\downarrow}-\mu(n_{i\uparrow}+n_{i\downarrow})+\frac{\omega_0}{2}(P_i^2+X_i^2)+\sqrt{2}gX_i n_{i\uparrow}n_{i\downarrow}\Big],
\label{H}
\end{equation}
where $v_{ij}$ denotes the hopping amplitude, $U$ the onsite repulsion, $\mu$ the chemical potential and $n_{i\sigma}$ measures the occupation of site $i$ with electrons of spin $\sigma$. The coupling of the double occupancy to a harmonic oscillator with frequency $\omega_0$ allows to describe the expansion of the atomic orbital (and associated reduction of the Coulomb interaction) which occurs when the orbital is occupied by two electrons. $X_i$ and $P_i$ are the boson position and momentum operators, respectively, and $g$ is the electron-boson coupling. (Here, the expansion of the orbital would correspond to a shift of $X$ in the negative direction.) In DMFT,\cite{Georges1996} this lattice model is mapped to a self-consistent solution of a quantum impurity model with Hamiltonian $H=H_\text{loc}+H_\text{hyb}+H_\text{bath}$ and a local term
\begin{equation}
H_\text{loc}=\underbrace{U d-\mu(n_{\uparrow}+n_{\downarrow})}_{\equiv H_\text{electron}}+\underbrace{({\omega_0}/{2})(P^2+X^2)}_{\equiv H_\text{boson}}+\sqrt{2}gX d,
\end{equation}
where $d=n_\uparrow n_\downarrow$ measures the double occupation. The hybridization term $H_\text{hyb}=\sum_{k,\sigma} (V_{k\sigma}a^\dagger_{k\sigma}c_\sigma+h.c.)$ and the noninteracting electron bath $H_\text{bath}=\sum_{k,\sigma} \epsilon_k a^\dagger_{k\sigma} a_{k\sigma}$ are determined by a self-consistent procedure in such a way that the bath mimics the lattice environment. In the action representation of the impurity model, the coupling to the environment appears as a hybridization term $S_\text{hyb}=-i\sum_{\sigma}\int dtdt'  \, c_{\sigma}^\dagger(t) \Delta_\sigma(t,t') c_{\sigma}(t') $, and the hybridization function $\Delta_\sigma(t,t')$ is determined by the DMFT self-consistency.  

To solve this impurity model we perform a unitary transformation\cite{Lang1962} which transforms operators $O\rightarrow \tilde O=e^{-iPX_0} O e^{iPX_0}$, and thus shifts $X\rightarrow X-X_0$. The transformed local Hamiltonian reads
\begin{align}
\tilde H_\text{loc} &= \tilde H_\text{electron}+H_\text{boson}
+\frac{\omega_0}{2}X_0^2-\sqrt{2}g d X_0 + X(\sqrt{2}gd-\omega_0X_0),
\label{trafo}
\end{align}
where we denote transformed operators by a tilde. 
The electrons and bosons can be decoupled by choosing $X_0=\frac{\sqrt{2}g}{\omega_0}d$, so that the unitary transformation becomes 
$\tilde O = W^\dagger O W$ with 
\begin{equation}
W=e^{iP\frac{\sqrt{2}g}{\omega_0}d},
\end{equation}
and the 
terms involving $X_0$ in Eq.~(\ref{trafo}) evaluate to $\frac{\omega_0}{2}X_0^2-\sqrt{2}gX_0d=-\frac{g^2}{\omega_0}d$. The transformed interaction and chemical potential therefore read
\begin{align}
\tilde U&=U-\frac{g^2}{\omega_0},\quad \tilde \mu=\mu.
\end{align}

We also have to evaluate the transformed electron creation and annihilation operators,
$\tilde c_\sigma^{(\dagger)} = W^\dagger c_\sigma^\dagger W = e^{-iP\frac{\sqrt{2}g}{\omega_0}d} c_\sigma^{(\dagger)} e^{iP\frac{\sqrt{2}g}{\omega_0}d}$. The creation operator gives a nonzero result only if it acts on a state with $n_\sigma=0$. In this case, the operator $W$ on the right becomes $1$, and the operator $W^\dagger$ on the left becomes $e^{-iP\frac{\sqrt{2}g}{\omega_0}n_{\bar\sigma}} $. Hence, we find $\tilde c_\sigma^\dagger = e^{-iP\frac{\sqrt{2}g}{\omega_0}n_{\bar\sigma}} c_\sigma^\dagger$, and similarly  $\tilde c_\sigma = e^{iP\frac{\sqrt{2}g}{\omega_0}n_{\bar\sigma}} c_\sigma$. Writing $P=\frac{i}{\sqrt{2}}(b^\dagger-b)$ in terms of the boson creation and annihilation operators and separating the contributions from the occupied and empty $\bar\sigma$ state, we finally get
\begin{align}
\tilde c_\sigma^\dagger &= e^{\frac{g}{\omega_0}(b^\dagger-b)}C^\dagger_\sigma+\bar C^\dagger_\sigma,\label{cdag}\\
\tilde c_\sigma &= e^{-\frac{g}{\omega_0}(b^\dagger-b)}C_\sigma+\bar C_\sigma,
\label{ctilde}
\end{align}
where we have introduced the Hubbard operators  
$C_\sigma^{(\dagger)} = n_{\bar\sigma}c_\sigma^{(\dagger)}$ and $\bar C_\sigma^{(\dagger)} =(1-n_{\bar\sigma})c_\sigma^{(\dagger)}$. Using Eqs.~\eqref{cdag} and \eqref{ctilde}, the hybridization term of the impurity action can be written as
$\tilde S_\text{hyb}\!=\!
-i\sum_{\sigma}\int dtdt'  \,  
\!
\Big[
e^{\frac{g}{\omega_0}(b^\dagger(t)-b(t))}C^\dagger_\sigma
(t)\Delta_\sigma(t,t')  
C_\sigma(t')
e^{\frac{g}{\omega_0}(b(t')-b^\dagger(t'))}
\!+
\bar C^\dagger_\sigma(t)
\Delta_\sigma(t,t')  
\bar C_\sigma(t')
+
... \Big]
$.
(The remaining terms involving products $C^\dagger \bar C $ and $\bar C^\dagger C $ 
are not written explicitly 
because they do not contribute to the lowest order expansion considered below.) 

\begin{figure}[t]
\begin{center}
\includegraphics[angle=0, width=0.55\columnwidth]{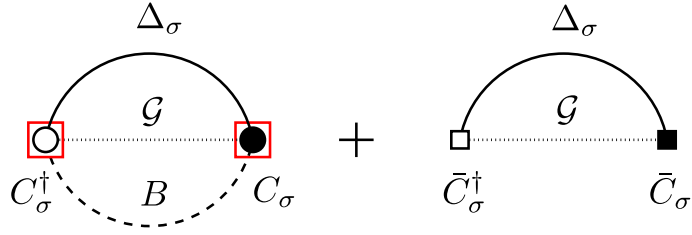}
\caption{
Illustration of the NCA self-energy diagrams for the dynamic Hubbard model. An open circle corresponds to an operator $C^\dagger_\sigma=n_{\bar\sigma}c_\sigma^\dagger$, a full circle to $C_\sigma=n_{\bar\sigma}c_\sigma$, an open square to $\bar C^\dagger_\sigma=(1-n_{\bar\sigma})c_\sigma^\dagger$ and a full square to $\bar C_\sigma=(1-n_{\bar\sigma})c_\sigma$. $\Delta_\sigma$ is the hybridization function, $\mathcal{G}$ a pseudo-particle Green's function, and $B$ the bosonic propagator defined in Eq.~(\ref{boson}). Red squares represent the ``box weight" (\ref{box}) in a simulation with externally driven bosons.
}
\label{fignca}
\end{center}
\end{figure}

As introduced for the Holstein-Hubbard model in Ref.~\onlinecite{Werner2013}, an approximation that captures the essential physics in the insulating phase and in the metal-insulator crossover regime at not too low temperature is to partially resum the diagrams of a hybridization expansion 
(non-crossing approximation, NCA),\cite{Keiter1971} and  to decouple the electron and boson terms by factorizing products  $e^{\frac{g}{\omega_0}(b^\dagger-b)}c^\dagger_\sigma (t) c_\sigma(t') e^{\frac{g}{\omega_0}(b-b^\dagger)}$  $\to$  $ c^\dagger_\sigma(t)  B(t,t') c_\sigma(t') $ in the action and in the Green function, with the bosonic function ($\mathcal{T}$ is the contour ordering operator) 
\begin{align}
B(t,t')&=-i\Big\langle 
\mathcal{T} 
e^{\frac{g}{\omega_0}(b^\dagger(t)-b(t))} e^{-\frac{g}{\omega_0}(b^\dagger(t')-b(t'))} \Big\rangle_{H_\text{boson}}\nonumber\\
&=-i\exp\Bigg[\frac{g^2/\omega_0}{\sinh(\beta\omega_0/2)}\Big(\cosh(\beta/2-i(t^>-t^<)\omega_0)-\cosh(\beta\omega_0/2)\Big)\Bigg].
\label{boson}
\end{align}
Here, $t^{</>}$ denotes the earlier/later time on the Keldysh contour.
The same electron-boson decoupling has recently been shown to provide the optimal representation of an Anderson impurity model with dynamically screened interactions in terms of a model with static interactions.\cite{Krivienko} In the case of the dynamic Hubbard model, we follow the same procedure and write 
$\tilde S_\text{hyb}
\to
-i\sum_{\sigma}\int dtdt'  \,  
\Big[
C^\dagger_\sigma
\Delta(t,t')B(t,t')  
C_\sigma
+
\bar C^\dagger_\sigma
\Delta(t,t')  
\bar C_\sigma
\Big].
$ 
Similarly, the contour-ordered Green function  $G_\sigma(t,t')=-i\langle 
\mathcal{T}
\tilde  c_\sigma(t) \tilde c_\sigma^\dagger(t')\rangle 
$ is factorized within the NCA like
\begin{align}
G_\sigma(t,t')
&=
-i\langle 
\mathcal{T}
C_\sigma(t)
C^\dagger_\sigma (t')
\rangle 
B(t',t)
-i\langle 
\mathcal{T}
\bar C_\sigma(t)
\bar C^\dagger_\sigma(t')
\rangle.
\end{align}
The correlation functions
$-i\langle 
\mathcal{T}
\bar C_\sigma(t)
\bar C^\dagger_\sigma(t')
\rangle
$
and 
$-i\langle 
\mathcal{T}
C_\sigma(t)
C^\dagger_\sigma(t')
\rangle
$
are then evaluated using the standard NCA diagrammatic expressions,\cite{Eckstein2010_nca} as illustrated in Fig.~\ref{fignca}. 

Because the dynamic Hubbard model is not particle-hole symmetric, an external driving of the bosons can be expected to yield nontrivial effects. Such a driving might have to be considered in the modeling of experiments similar to those recently reported by Kaiser {\it et al.},\cite{Kaiser2014} or Singla {\it et al.},\cite{Singla2014} where the lattice and hence the Wannier orbitals are shaken. This effect can be described within the present formalism, starting from the model
\begin{equation}
H_\text{loc}=H_\text{electron}+H_\text{boson} 
+(\sqrt{2}gd+f(t))X,
\end{equation}
with $f(t)$ some time-dependent function. Replacing $\sqrt{2}gd\rightarrow \sqrt{2}gd+f(t)$ in Eq.~(\ref{trafo}), the Lang-Firsov shift becomes $X_0=\frac{1}{\omega_0}(\sqrt{2}gd+f(t))$, and the shifts of the interaction and chemical potential are obtained from $\frac{\omega_0}{2}X_0^2-(\sqrt{2}gd+f)X_0=-\frac{\omega_0}{2}X_0^2=-\frac{1}{\omega_0}((g^2+\sqrt{2}gf)d+\frac{1}{2}f^2)$. Hence 
\begin{equation}
\tilde U(t)=U-\frac{1}{\omega_0}(g^2+\sqrt{2}gf(t)), \label{utilde}
\end{equation}
$\tilde \mu=\mu$, and there is an additional time-dependent energy shift $-\frac{1}{2\omega_0}f^2(t)$. 

According to the discussion in Section IIB of Ref.~\onlinecite{Werner2013}, the time-dependent unitary transformation $W(t)$ implies an additional term $i\dot W(t)^\dagger W(t)$ in the Hamiltonian. Including this term, the bosonic part of the transformed Hamiltonian becomes 
\begin{equation}
\tilde H_\text{boson}=\frac{\omega_0}{2}(P^2+X^2)+\frac{1}{\omega_0}f'(t)P.
\end{equation}
From Eq.~(\ref{cdag}) and the solution of the Heisenberg equations of motion for the $b$-operators, $b^\dagger(t)=b^\dagger e^{i\omega_0t}+\int_0^t d\bar t \frac{1}{\sqrt{2}\omega_0}f'(\bar t)e^{i\omega_0(t-\bar t)}$, one finds (assuming $f(0)=0$)
\begin{equation}
\tilde c^\dagger_\sigma(t)=e^{\frac{g}{\omega_0}(b^\dagger e^{i\omega_0t}-be^{-i\omega_0t})}e^{\frac{g}{\omega_0}\sqrt{2}i\int_0^td\bar t f(\bar t)\cos(\omega_0(t-\bar t))}C^\dagger_\sigma+\bar C^\dagger_\sigma.
\end{equation}
In the hybridization expansion, all vertices $C_\sigma$ and $C_\sigma^\dagger$ are thus multiplied by the ``box weight" 
\begin{equation}
e^{s\frac{g}{\omega_0}\sqrt{2}i\int_0^td\bar t f(\bar t)\cos(\omega_0(t-\bar t))}, \label{box}
\end{equation}
with $s=\pm 1$ for creation and annihilation operators (see Fig.~\ref{fignca}). 
If the driving is of the form $f(t)=A\sin(\omega t)$, then 
$\int_0^td\bar t f(\bar t)\cos(\omega_0(t-\bar t))=\frac{A\omega}{\omega^2-\omega_0^2}(\cos(\omega_0t)-\cos(\omega t))$ for $\omega\ne\omega_0$, and $\frac{At}{2}\sin(\omega_0 t)$ for $\omega=\omega_0$.

\section{Results}
\label{results}
\subsection{``Photo-doping" of the Mott insulator}

Since the approximate impurity solver described in the previous section is expected to give qualitatively correct results in the insulating phases of the model, we will consider a ``photo-doping" type perturbation in a paramagnetic Mott insulator. For simplicity, we do not simulate the excitation of carriers across the Mott-gap by an electric field pulse, but instead apply a short $U$-pulse to the system: 
\begin{equation}
U(t)=\left\{
\begin{array}{ll}
U+\Delta U \sin(\omega_U t) \quad & 0\le t \le 4\pi/\omega_U,\\
U \quad & t> 4\pi/\omega_U.
\end{array}
\right.
\label{upulse}
\end{equation}
For $\omega_U=U$, the effect of this modulation is to promptly create a certain number of doublon-hole pairs, so that  the state of the system after the $U$-pulse is similar to a photo-doped state after impulsive stimulation with a laser pulse with frequency comparable to $U$. We assume a semi-circular density of states of bandwidth $4v$, so that the DMFT self-consistency loop simplifies to $\Delta_\sigma(t,t')=v^2 G_\sigma(t,t')$,\cite{Aoki2014} and we use $v$ as the unit of energy.   

The photo-doping induced changes in the spectral function of the (particle-hole symmetric) Holstein-Hubbard model have been investigated in Ref.~\onlinecite{Werner2015}. In contrast to the Hubbard model, where the Mott insulating spectral function is little affected by photo-doping,\cite{Eckstein2011} 
in particular in the gap region, the rapid cooling of the photo-doped carriers and the possibility of polaron formation in the Holstein-Hubbard model with strong electron-phonon coupling can lead to a series of doping-induced in-gap states. The appearance of these polaronic states also has an important effect on the life-time of the photo-carriers, which becomes much shorter than in the Hubbard model, and also doping dependent ($d/dt\langle d\rangle \propto (\langle d\rangle-d_\text{th})^2$, with $d_\text{th}$ the doublon population of the initial state).

\begin{figure}[t]
\begin{center}
\includegraphics[angle=-90, width=0.32\columnwidth]{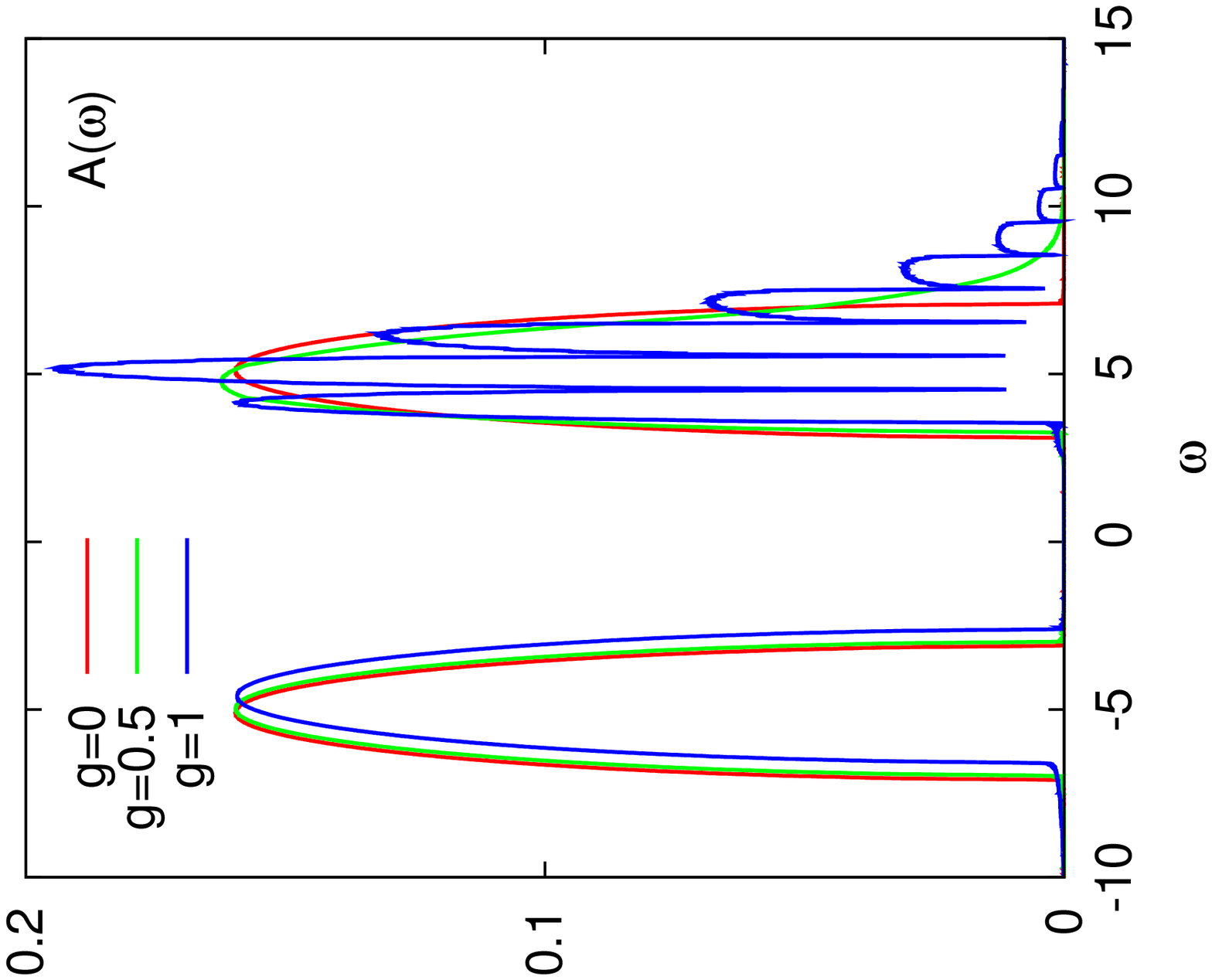}\hfill
\includegraphics[angle=-90, width=0.32\columnwidth]{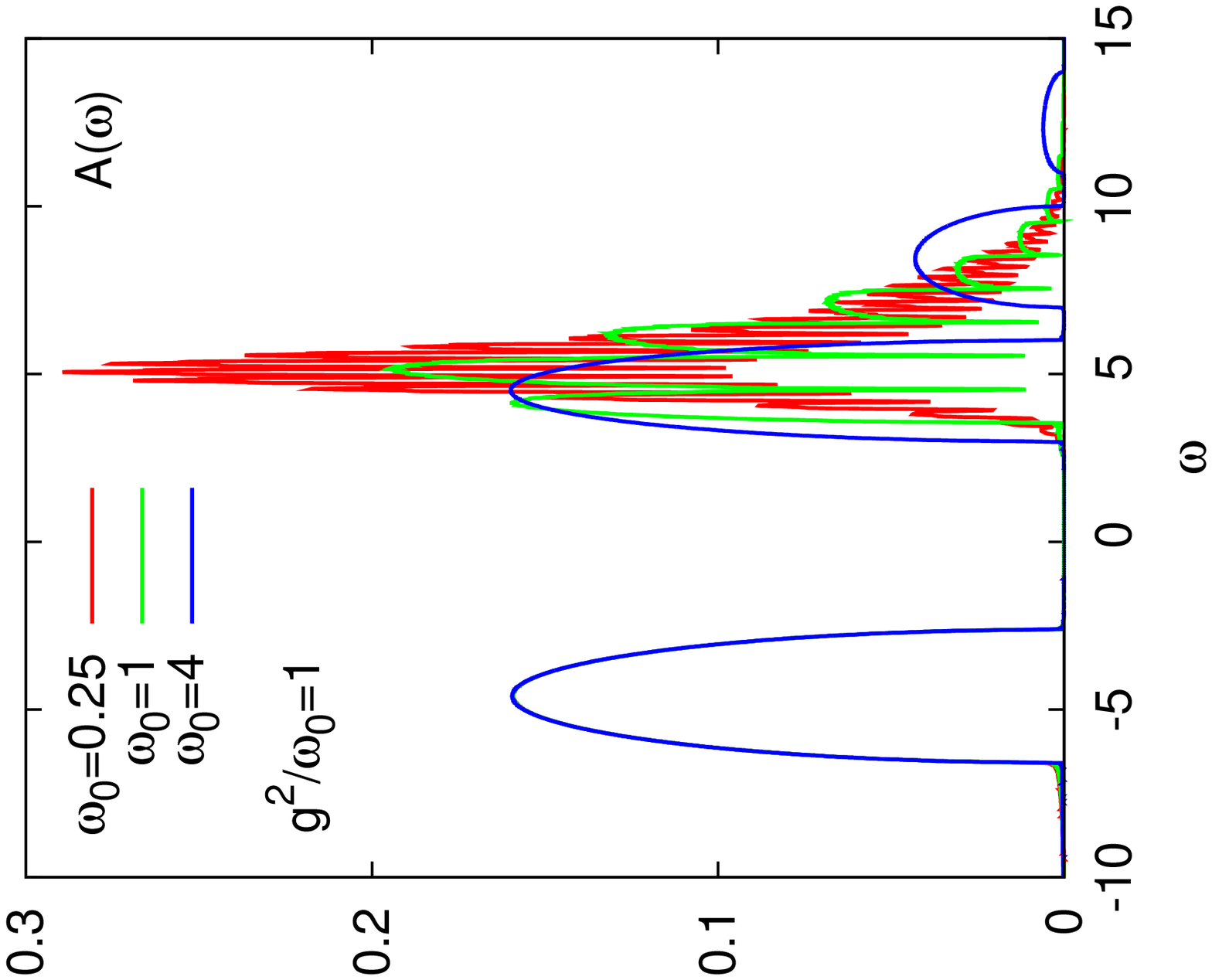}\hfill
\includegraphics[angle=-90, width=0.32\columnwidth]{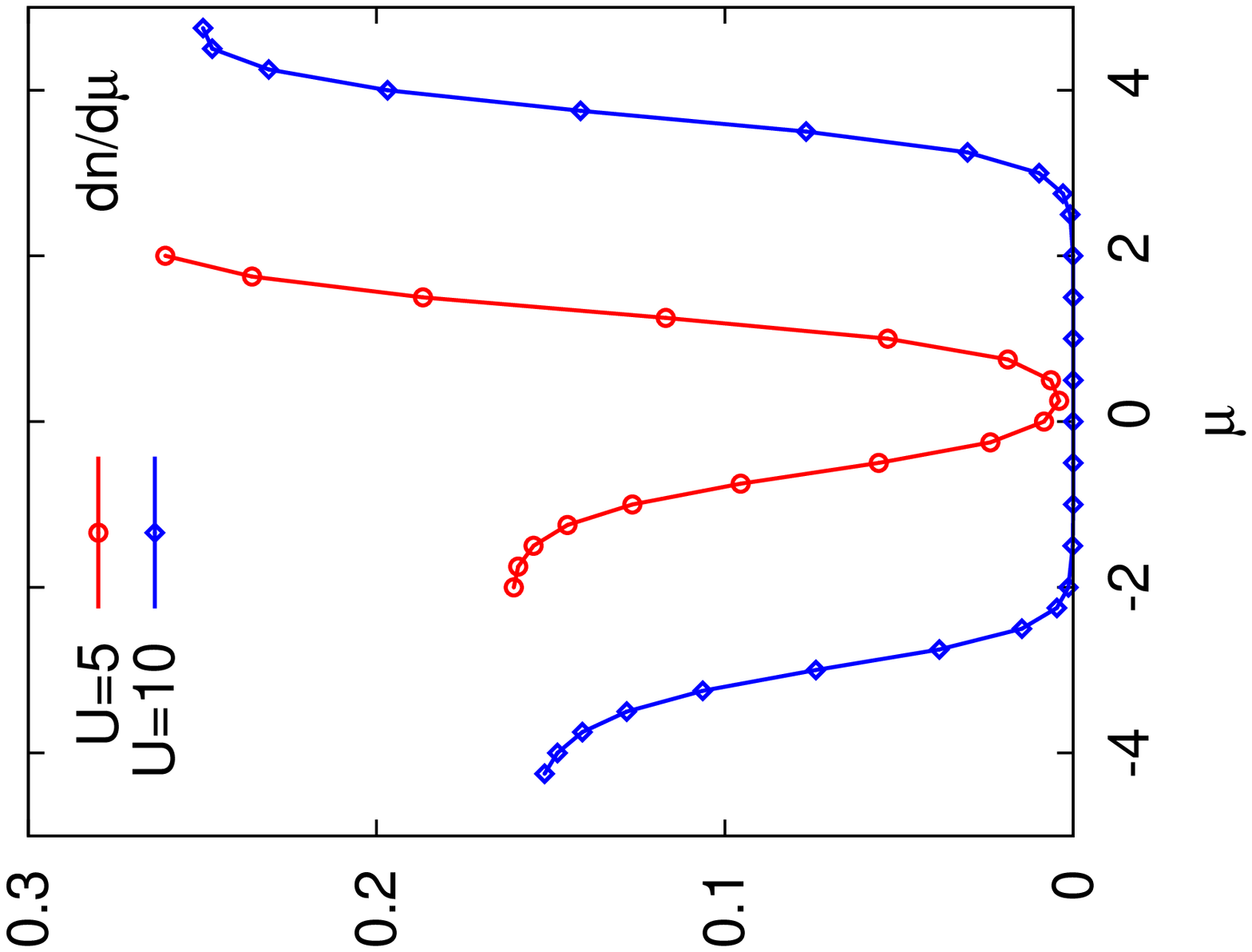}
\caption{Equilibrium spectral functions and compressibilities in the dynamic Hubbard model. The left panel shows spectral functions for $U=10$, $\mu=0$, $\omega_0=1$ and indicated values of $g$, while the middle panel shows spectral functions for $U=10$, $\mu=0$, fixed $g^2/\omega_0=1$ and indicated values of $\omega_0$. The right panel plots the compressibility as a function of $\mu$ for $g=1$, $\omega_0=1$, $U=5$ and $U=10$. The inverse temperature is $\beta=5$ for all plots.
}
\label{fig_eq}
\end{center}
\end{figure}

\subsection{Equilibrium results}

We now turn to the dynamic Hubbard model (\ref{H}) and first briefly discuss the spectral properties of the Mott insulating equilibrium system. The left panel of Fig.~\ref{fig_eq} plots the spectral function 
$A(\omega)$ for $U=10$, $\mu=0$, $\omega_0=1$ and indicated values of $g$. The inverse temperature is $\beta=5$. Because the chemical potential is inside a large gap, the filling is approximately $n=1$ independent of $g$. For $g=0$ (Hubbard model), the system is particle-hole symmetric and the spectral function features two Hubbard bands of approximately semi-circular shape. As the electron-boson coupling is increased, the spectral function becomes asymmetric: while the lower Hubbard band essentially retains its semi-circular shape, the upper Hubbard band broadens and for large enough $g$ splits into sidebands. In addition, there is a small shift of the Hubbard bands to higher $\omega$. If the parameter $g^2/\omega_0>0$ is kept fixed while $\omega_0$ is varied, the separation between the sidebands changes, while the broadening of the upper Hubbard bands stays roughly the same (middle panel).

The boson-induced broadening of the upper Hubbard band in the small-$g$ regime is a manifestation of the reduced doublon correlations, which is the effect described in the original work by Hirsch. The splitting into sidebands at larger $g$ is due to the formation of polarons. We will not discuss here under which conditions this polaronic regime may be realized, but simply explore the nonequilibrium properties of the dynamic Hubbard model in the full range of boson couplings.

One can also observe the asymmetry in the electron and hole correlations by computing the density $n(\mu)$ or the compressibility $dn/d\mu$. As shown in the right panel of Fig.~\ref{fig_eq}, the compressibility is larger on the electron-doped side than on the hole-doped side, which is consistent with weaker doublon correlations.

\subsection{Spectral functions of the photo-doped system}

\begin{figure}[t]
\begin{center}
\includegraphics[angle=-90, width=0.49\columnwidth]{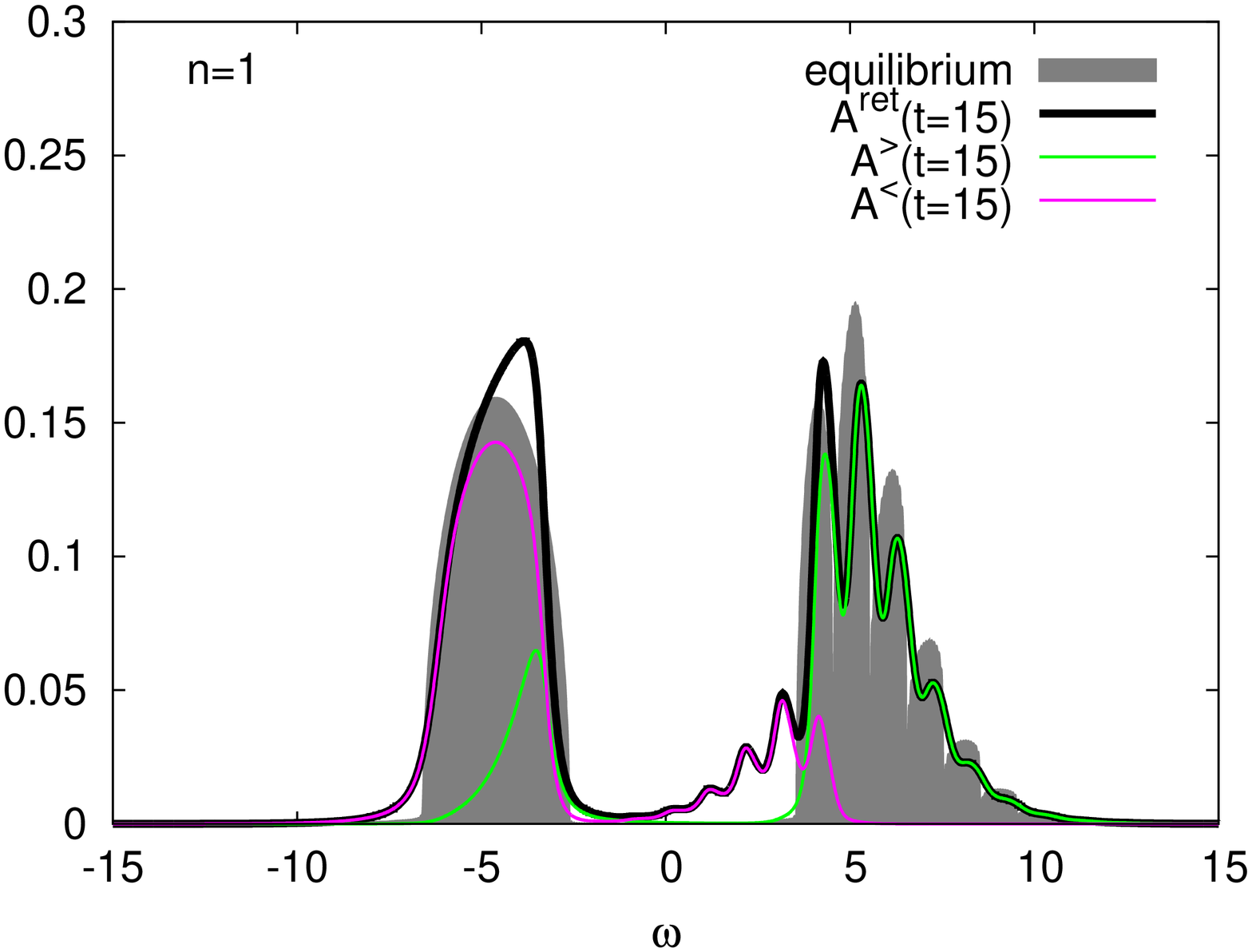}
\includegraphics[angle=-90, width=0.49\columnwidth]{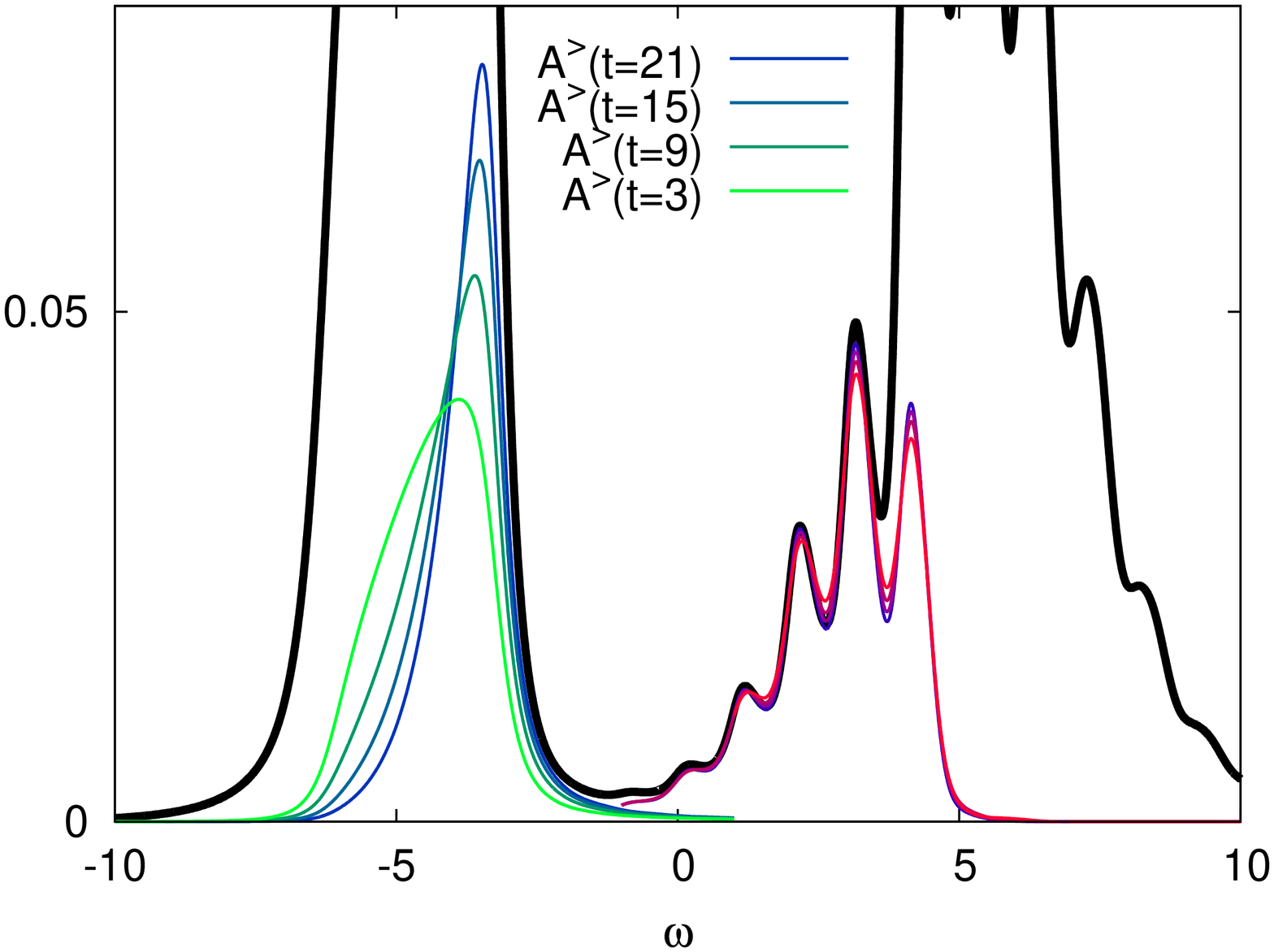}
\includegraphics[angle=-90, width=0.49\columnwidth]{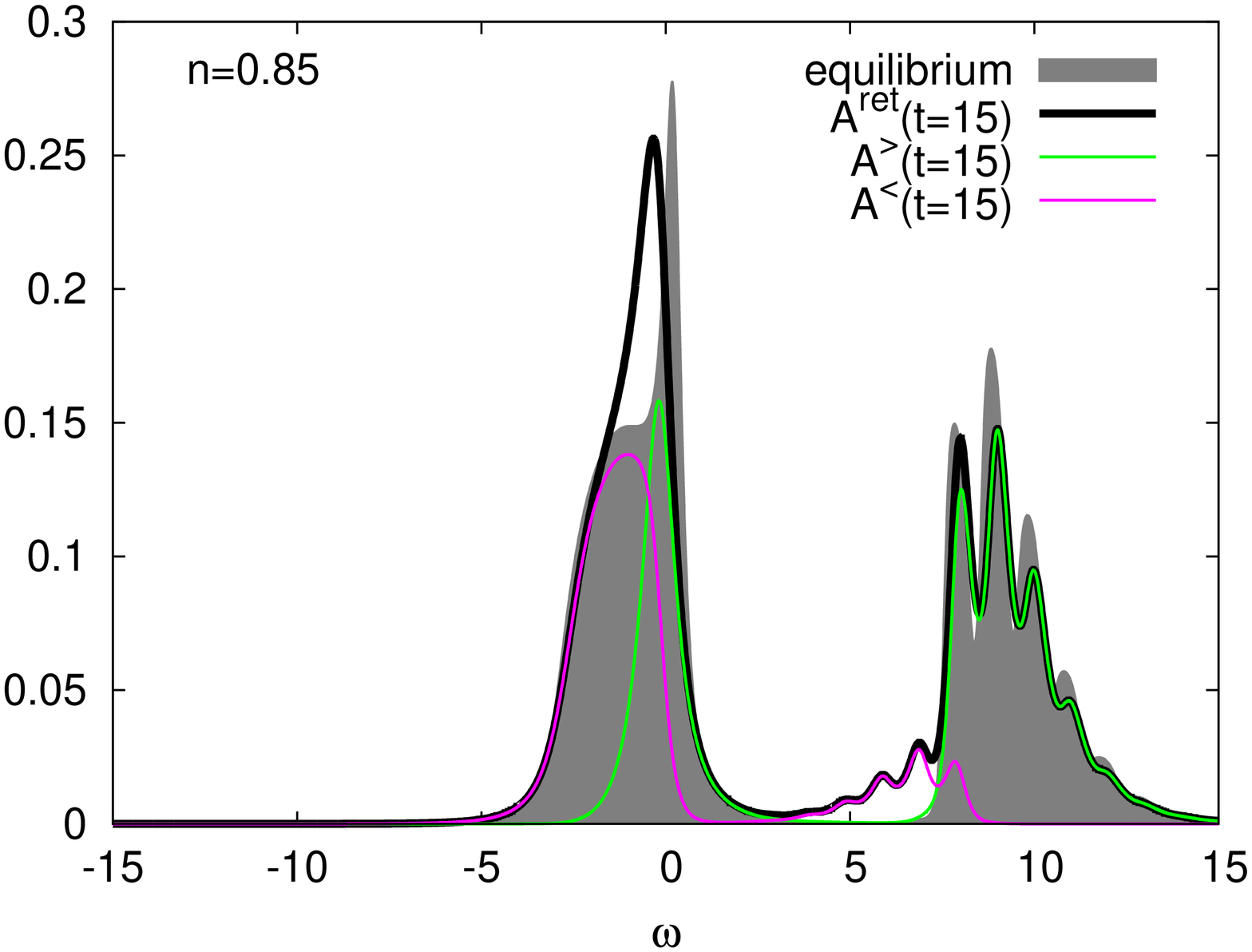}
\includegraphics[angle=-90, width=0.49\columnwidth]{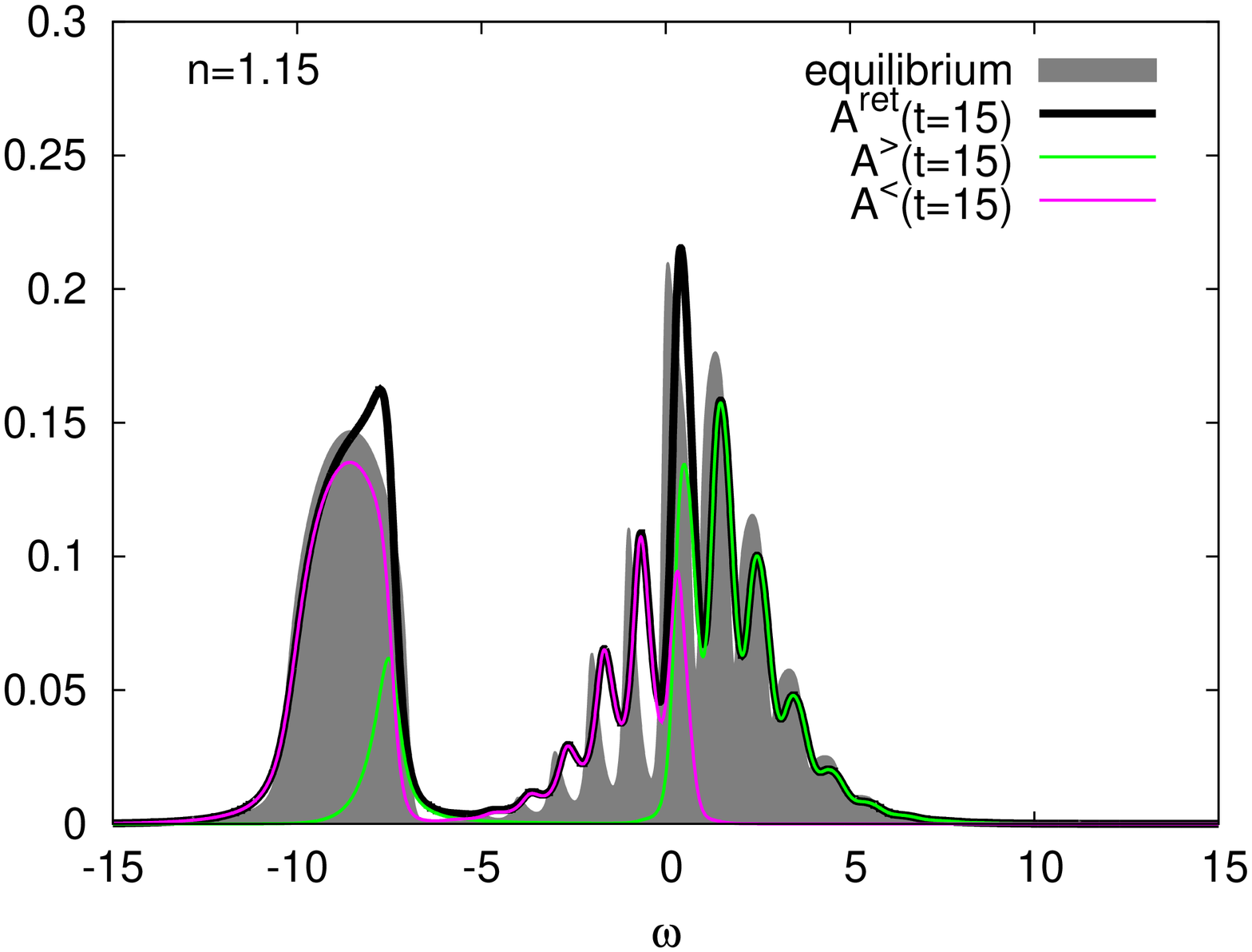}
\caption{Left and bottom panels: Nonequilibrium spectral functions measured at time $t=15$ in a half-filled, hole-doped, and electron-doped system after photo-doping. The parameters are $U=10$, $g=1$, $\omega_0=1$ and $\beta=5$. The frequency of the two-cycle excitation pulse Eq.~\eqref{upulse} is $\omega_U=U$ and the amplitude is $\Delta U=U$.  The black curve shows the total spectral function, while the green and pink curves indicate the distributions of occupied and empty states. For comparison, we also plot the initial equilibrium spectral functions. The top right panel shows the time evolution of the distribution functions for $n=1$.  
}
\label{photodoping}
\end{center}
\end{figure}

We next investigate the changes in the spectral function after the impulsive excitation defined in Eq.~\eqref{upulse}. Figure~\ref{photodoping} shows results for $U=10$, $g=1$, $\omega_0=1$ and pulse parameters $\Delta U=U$ and $\omega_U=U$. The top left panel corresponds to a half-filled system. We plot both the total spectral function $A^\text{ret}=A^<+A^>$, the distribution of occupied states $A^<$, and the distribution of unoccupied states $A^>$. These time-dependent spectra were obtained by forward time integration, using the formulas
\begin{align}
A^<(\omega,t)&=\frac{1}{\pi}\text{Im}\int_t^\infty dt' e^{i\omega(t'-t)}G^<(t',t),\\
A^\text{ret}(\omega,t)&=-\frac{1}{\pi}\text{Im}\int_t^\infty dt' e^{i\omega(t'-t)}G^\text{ret}(t',t).
\end{align}
We plot the results for $t=15$, i.e., during the relaxation which follows the short 2-cycle pulse (which lasts until $4\pi/\omega_U=1.26$). As can be seen by comparing the transient spectral function to the initial equilibrium result in grey, the photo-doping induces substantial changes, especially in the gap region. In analogy to the previous results for the Holstein-Hubbard model~\cite{Werner2015} we find that the production of doublons results in a series of in-gap states, which can be viewed as additional boson sidebands of the upper Hubbard band. However, in contrast to the particle-hole symmetric Holstein-Hubbard case, the sidebands are limited to the region $\omega>0$, and there are no in-gap states appearing in the vicinity of the lower Hubbard band. The lower Hubbard band merely exhibits an asymmetry reminiscent of a smeared-out quasi-particle band near the upper band edge.  

The distribution of occupied states shows that the additional doublons created by the pulse are rapidly cooled down and are the origin of the polaronic in-gap states. At the same time, the distribution of unoccupied states shows an accumulation of holes near the upper edge of the lower Hubbard band.  
The time-evolution of the doublon and hole populations for $n=1$ is illustrated in the top right panel. The curves with colors ranging from green to blue show the function $A^<$ in the lower Hubbard band from $t=3$ to $21$, and the set of curves with colors ranging from red to dark-violette show the $A^<$ function in the upper Hubbard band for the same times. One notices that the doublon population exhibits almost no time dependence, which shows that the cooling of the doublons and the appearance and population of the polaronic in-gap states is so fast that it essentially takes place during the application of the pulse. This observation is consistent with the Holstein-Hubbard results in Ref.~\onlinecite{Werner2015} for the same parameter regime. The redistribution of the hole population is a much slower process, which is expected because the holes do not couple directly to the bosons in the dynamic Hubbard model. Nevertheless, the cooling of the holes which is evident in this set of data is very fast compared to the relaxation in the Hubbard model ($g=0$), where there would be almost no change in the energy distribution of doublons and holes on the timescales of this plot.\cite{Eckstein2011} Apparently, the scattering with ``cold" doublons provides a relatively efficient dissipation mechanism for the holes. In the next section, we will analyze this relaxation process more quantitatively. 

We can repeat similar calculations also for hole and electron doped systems. In the bottom panels of Fig.~\ref{photodoping} we show results for $n=0.85$ and $n=1.15$. In the hole-doped system, the spectral function features a narrow quasi-particle peak at the Fermi energy. After the pulse, this peak broadens due to the effects of heating and photo-doping. Near the upper Hubbard band, we again observe the appearance of a series of polaronic side-bands. They are less prominent than in the half-filled case, because in the doped system, the doublon-hole recombination is faster, so that the number of photo-carriers in the upper Hubbard band is substantially reduced by $t=15$. In the electron-doped system, we observe a series of doping-induced side-bands already in the equilibrium state. After the pulse, the weight of these polaronic states is further increased, while in the lower Hubbard band we observe the appearance of a small ``quasi-particle" peak. Again, because of the faster doublon-hole recombination, the changes in the lower Hubbard band are smaller than in the half-filled case.     

\begin{figure}[t]
\begin{center}
\includegraphics[angle=-90, width=0.49\columnwidth]{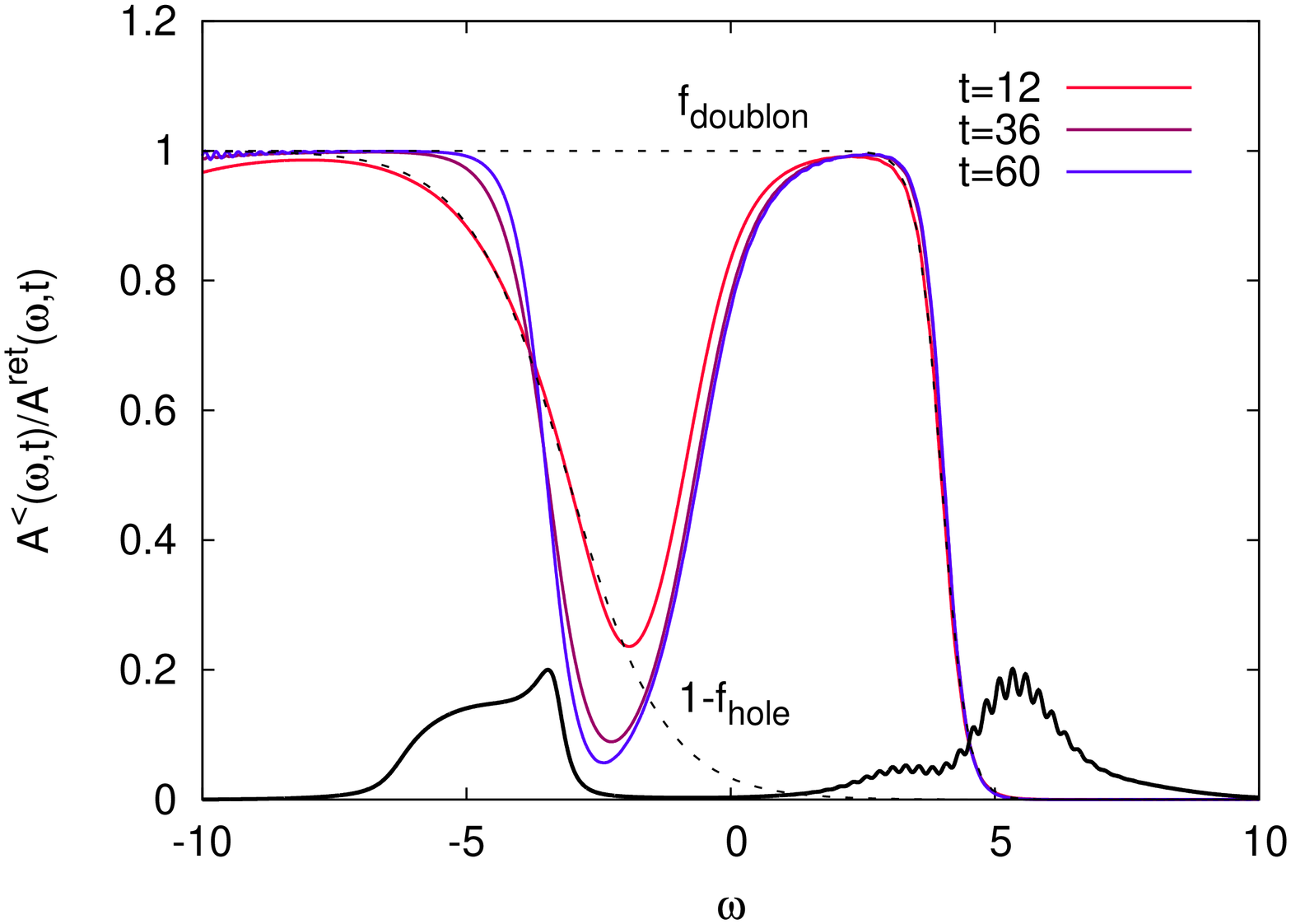}
\includegraphics[angle=-90, width=0.49\columnwidth]{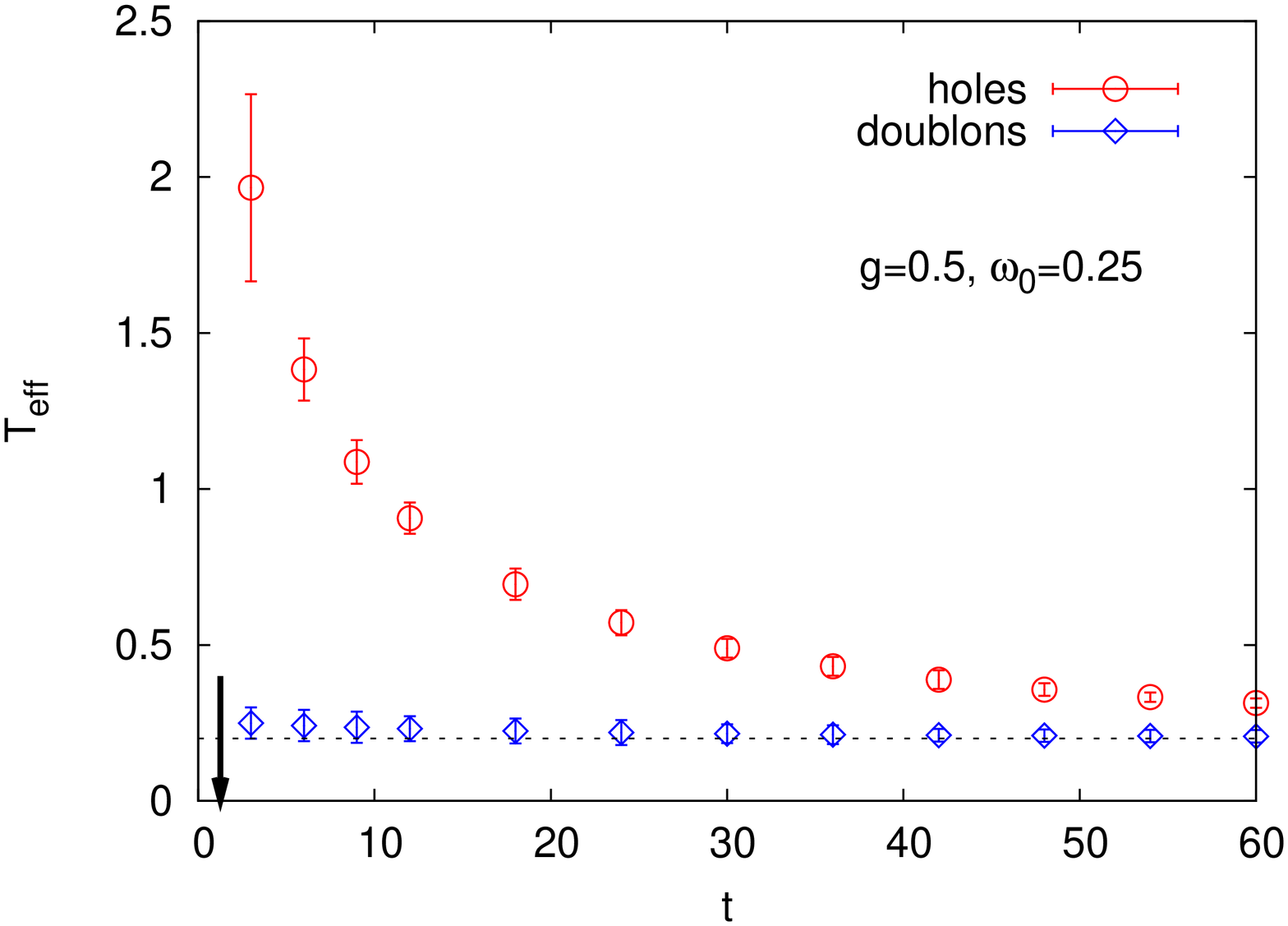}
\includegraphics[angle=-90, width=0.49\columnwidth]{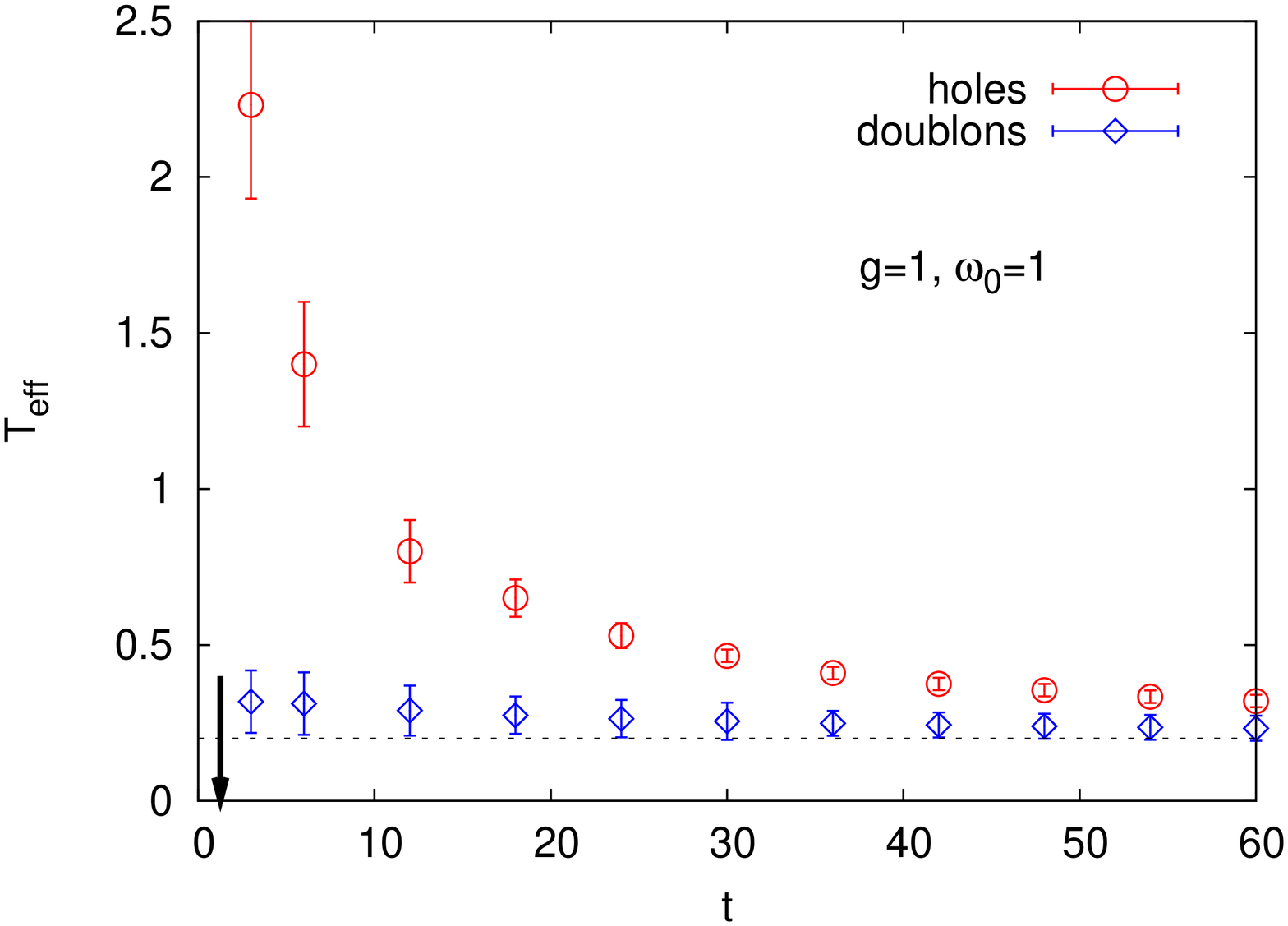}
\includegraphics[angle=-90, width=0.49\columnwidth]{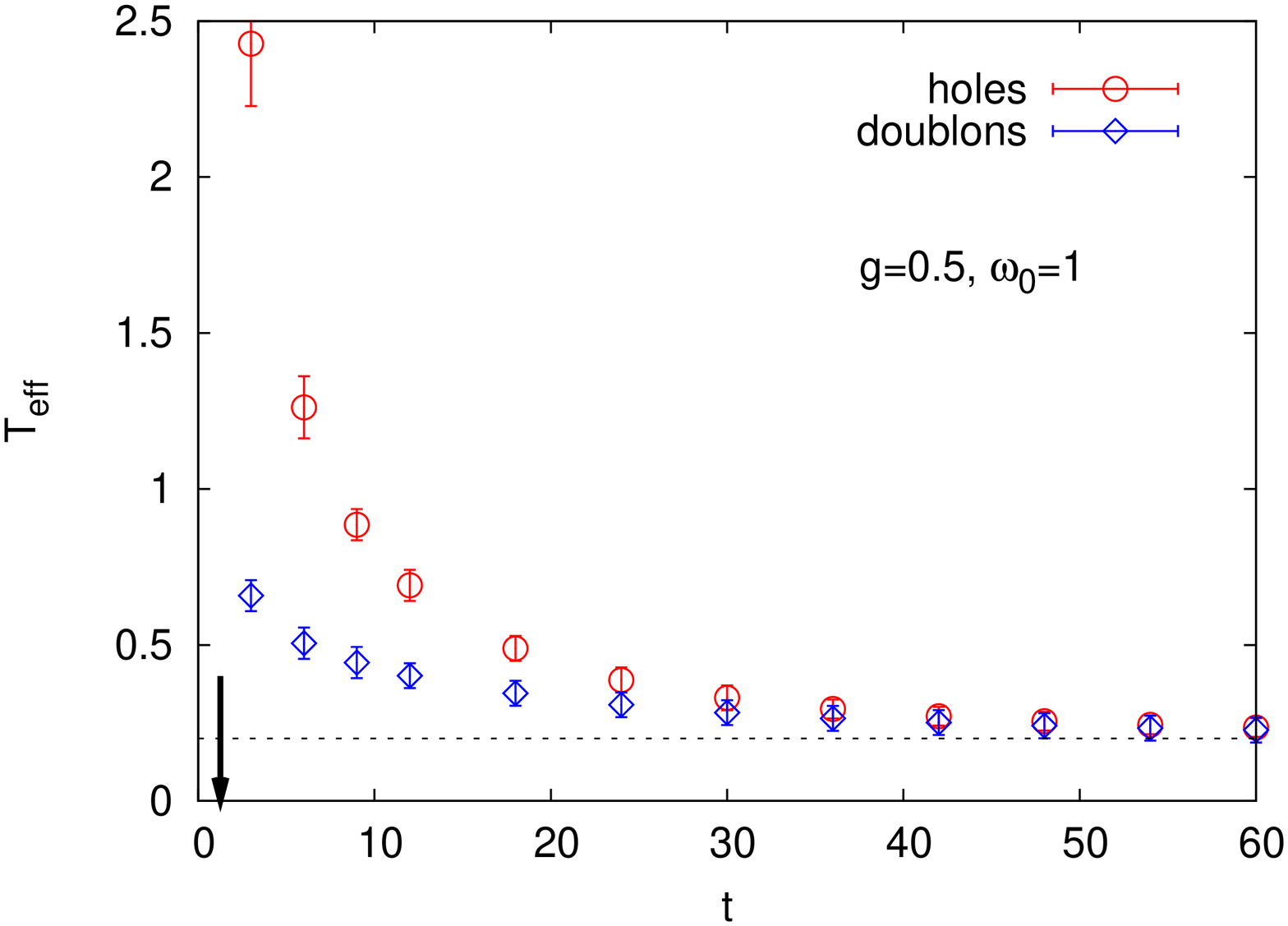}
\caption{Top left panel: Nonequilibrium distribution function $f(\omega,t)=A^<(\omega,t)/A^\text{ret}(\omega,t)$ for the model with $g=0.5$, $\omega_0=0.25$ and indicated values of $t$. To illustrate the energy range covered by the Hubbard bands, we also plot $A^\text{tot}(\omega,t=60)$ (black line). For the $t=12$ curve, we show the fits to Fermi distribution functions by dashed lines (fitting range $[-7,-2]$ and $[2,7]$). Top right panel: Effective electron and hole temperatures extracted from the Fermi function fits for $g=0.5$, $\omega_0=0.25$. The dashed line indicates the temperature of the initial state, and the arrow the end of the 2-cycle pulse. Bottom panels: Effective doublon and hole temperatures for different boson couplings and boson frequencies. 
}
\label{cooling}
\end{center}
\end{figure}

\subsection{Effective doublon and hole temperatures}

Except deep in the polaronic regime, the doublon and hole distribution functions quickly reach an approximately thermal shape in the energy region corresponding to the upper and lower Hubbard band. It is therefore instructive to analyze the relaxation process in terms of the effective ``temperatures" of the doublons and holes. To measure these temperatures, we compute the nonequilibrium distribution function $f(\omega,t)=A^{<}(\omega,t)/A^\text{ret}(\omega,t)$. In equilibrium, $f(\omega)$ corresponds to the Fermi function for the corresponding temperature. As shown in the top left panel of Fig.~\ref{cooling} for the model with $g=0.5$ and $\omega_0=0.25$, the photo-doped system does not have a Fermi like nonequilibrium distribution. Rather, there are two Fermi edges, one in the energy range of the upper Hubbard band ($\omega\approx 4$ in the figure), and one in the energy range of the lower Hubbard band ($\omega\approx -4$). By fitting the distribution functions to a Fermi function in these energy regions, one can extract an effective doublon and hole temperature, and in principle also the corresponding effective chemical potentials. For the earliest time ($t=12$) these fits are indicated in the figure by dashed black lines. From these fits, one extracts $T_\text{eff}^\text{hole}(t=12)\approx 0.23$ and $T_\text{eff}^\text{hole}(t=12)\approx 0.92$. 

The time evolution of these effective electron and hole temperatures is illustrated in the top right panel. Since the Fermi fits depend on the fitting range, there is some uncertainty associated with the calculation of the effective temperatures, and the error bars in the figure give a conservative estimate of this uncertainty. It is obvious that the doublons, which couple directly to the bosons, are rapidly cooled down to an effective temperature close to the initial equilibrium temperature of $1/\beta=0.2$ (horizontal dashed line), while the holes are still in the process of dissipating their energy at the longest accessible time. We note that in our closed system, the energy injected by the pulse will lead to a thermalization at a temperature above the initial equilibrium temperature, so that the effective temperature of the doublons might increase on longer timescales. 
However, since the energy changes associated with the redistribution of spectral weight within the Hubbard bands and the cooling of doublons and holes during the initial stage of the relaxation can apparently be easily absorbed by the bosons, such an increase of the doublon temperature would be controlled by the doublon-hole recombination time, which is very long in large-gap insulators.\cite{Eckstein2011,Sensarma2010}

The bottom panels show similar results for different boson couplings and boson frequencies. Obviously, the cooling rates and the ratio between effective doublon and hole temperatures depends on these parameters in a nontrivial way.

\subsection{Dependence of the cooling rate on the electron-electron and electron-boson interaction}

\begin{figure}[t]
\begin{center}
\includegraphics[angle=-90, width=0.32\columnwidth]{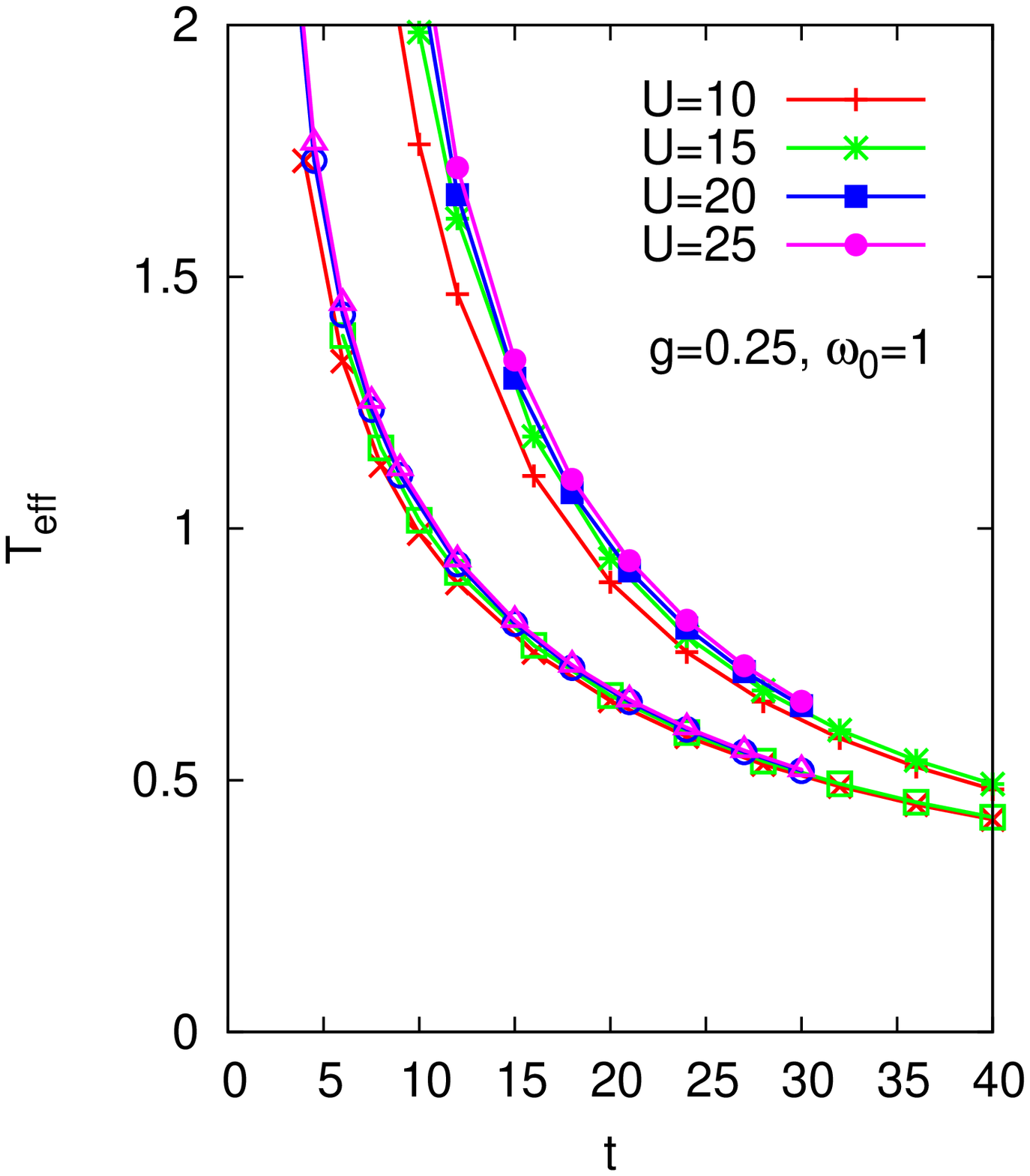}\hfill
\includegraphics[angle=-90, width=0.32\columnwidth]{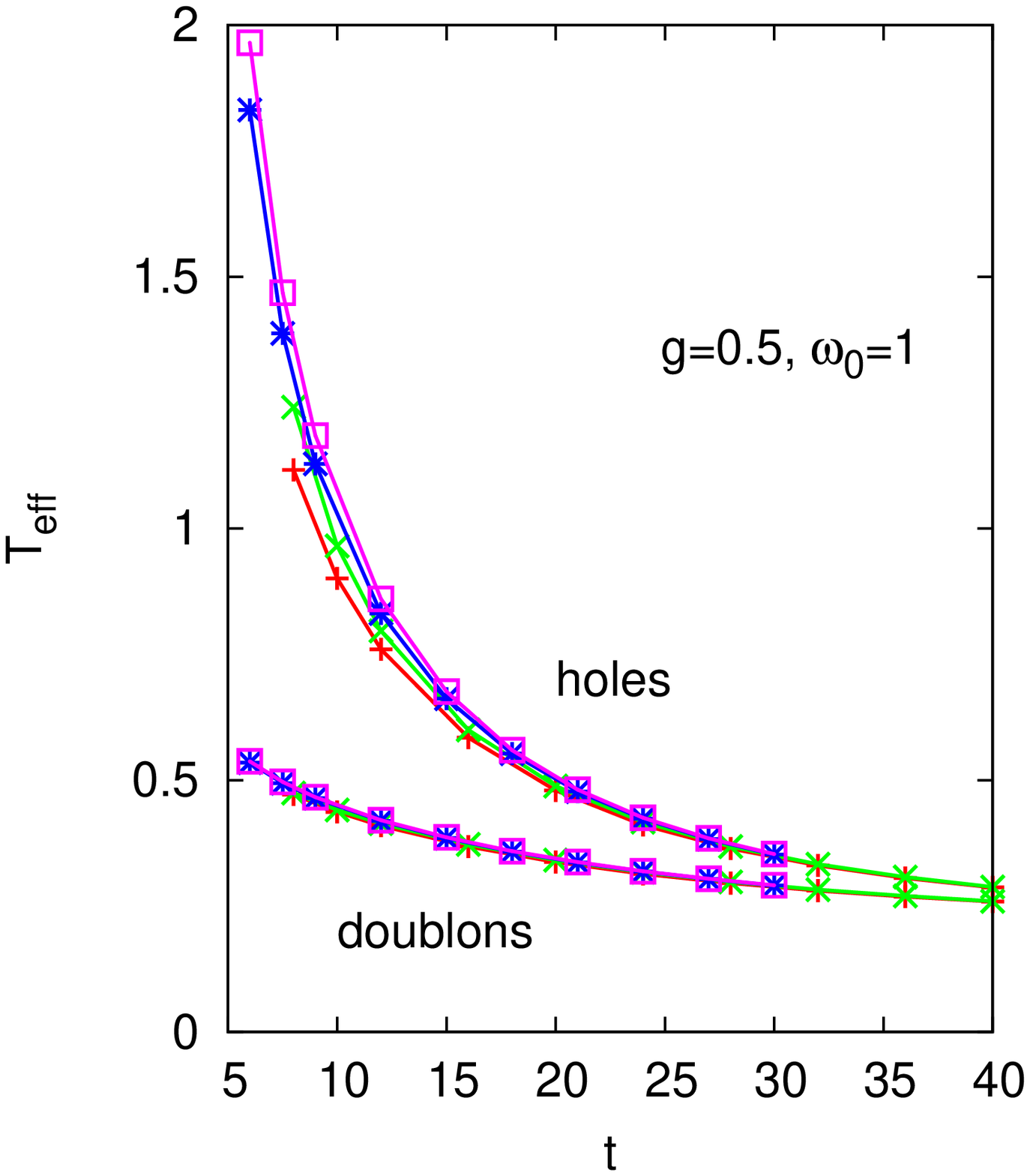}\hfill
\includegraphics[angle=-90, width=0.32\columnwidth]{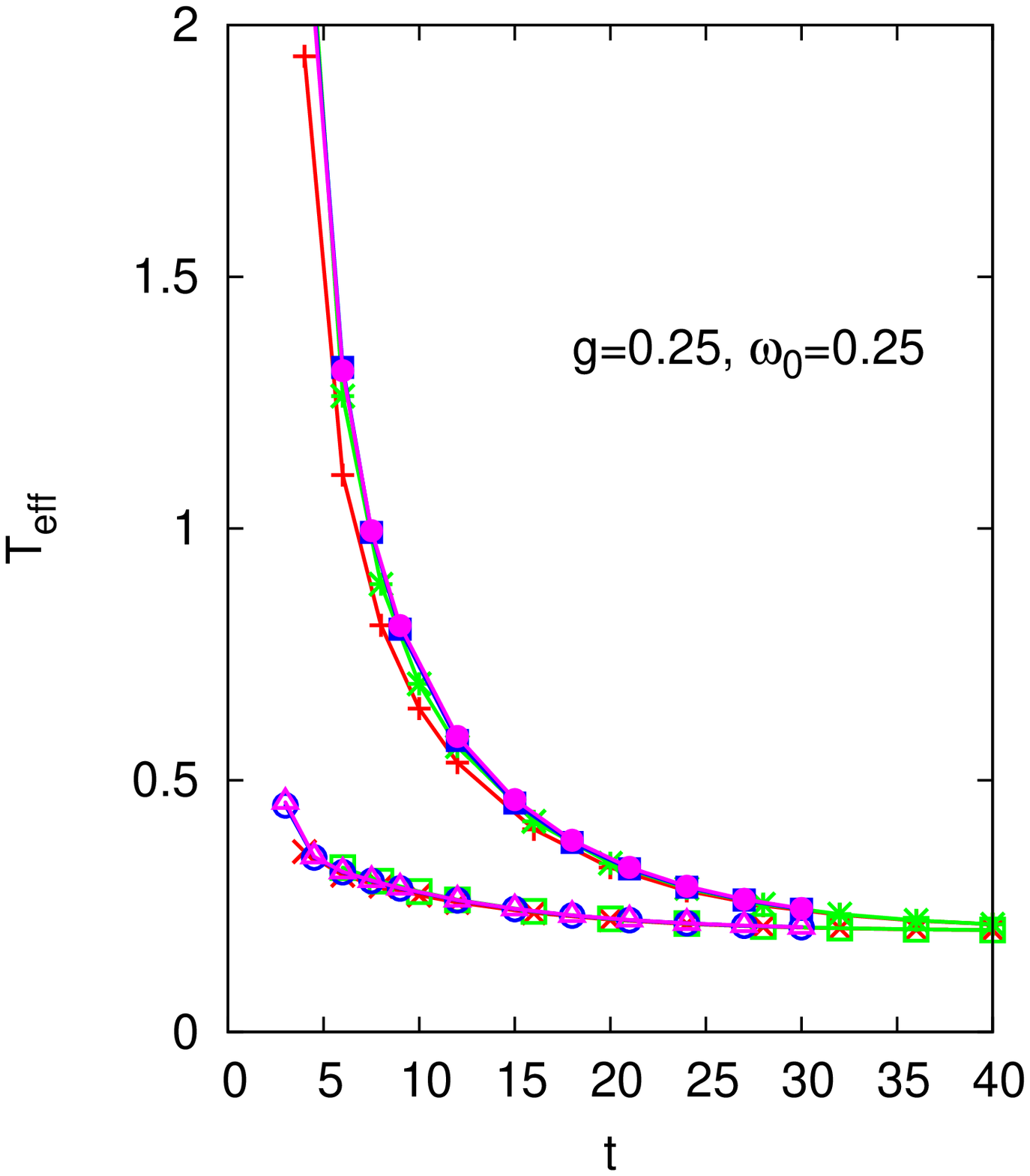}
\caption{Effective doublon and hole temperatures in systems with photo-doping concentration $0.1$, indicated boson couplings and boson energies, and different values of $U$. The upper curves show the hole temperatures and the lower curves the doublon temperatures. 
}
\label{cooling_U}
\end{center}
\end{figure}

While the coupling to the bosons provides a dissipation mechanism for the doublons, the cooling of the holes is an indirect process which involves doublon-hole scattering. Here, one could imagine scattering processes with virtual break-up of the doublon, $d+h \rightarrow\,\, \uparrow+\downarrow \,\,\rightarrow d+h$, or the scattering between a high energy hole and a low-energy doublon, which transfers energy from the hole to the doublon without such virtual excitations. If the former processes are relevant, one should observe a dependence of the cooling rate on $U$, because the local energies of the states $d+h$ and $\uparrow+\downarrow$ differ by $U$. Figure \ref{cooling_U} shows the $U$-dependence of the effective doublon and hole temperatures for different boson couplings and boson energies. There is a small dependence of the effective hole temperature on $U$, which suggests a more rapid cooling in small-gap insulators, and thus a certain role of the first type of scattering processes in the initial stage of the relaxation. The effective temperatures at later times are however essentially independent of $U$. It thus appears that the dominant scattering mechanism is of the second type, i.e., energy transfer from hot holes to cold doublons without virtual excitations.    

\begin{figure}[t]
\begin{center}
\includegraphics[angle=-90, width=0.32\columnwidth]{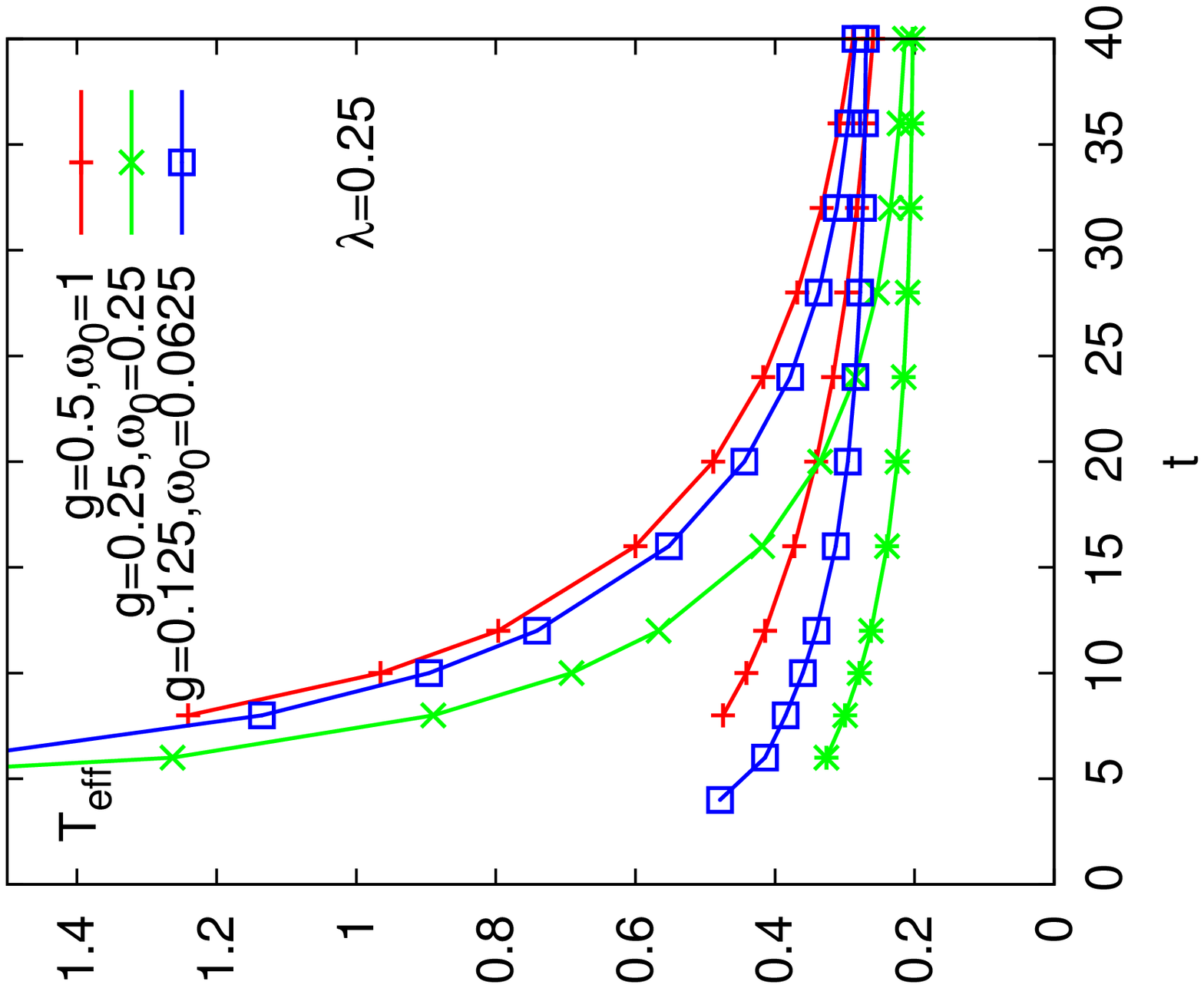}\hfill
\includegraphics[angle=-90, width=0.32\columnwidth]{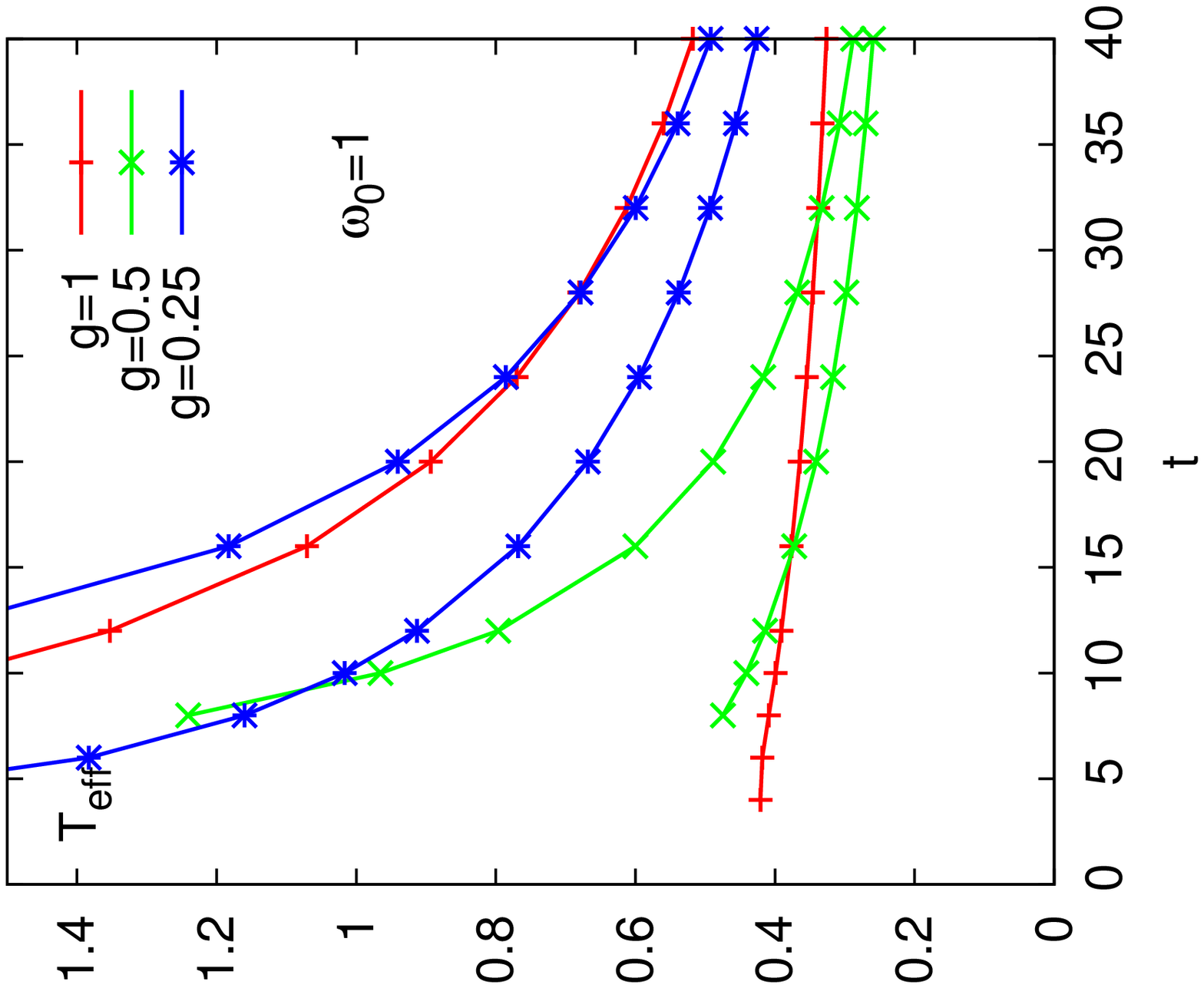}\hfill
\includegraphics[angle=-90, width=0.32\columnwidth]{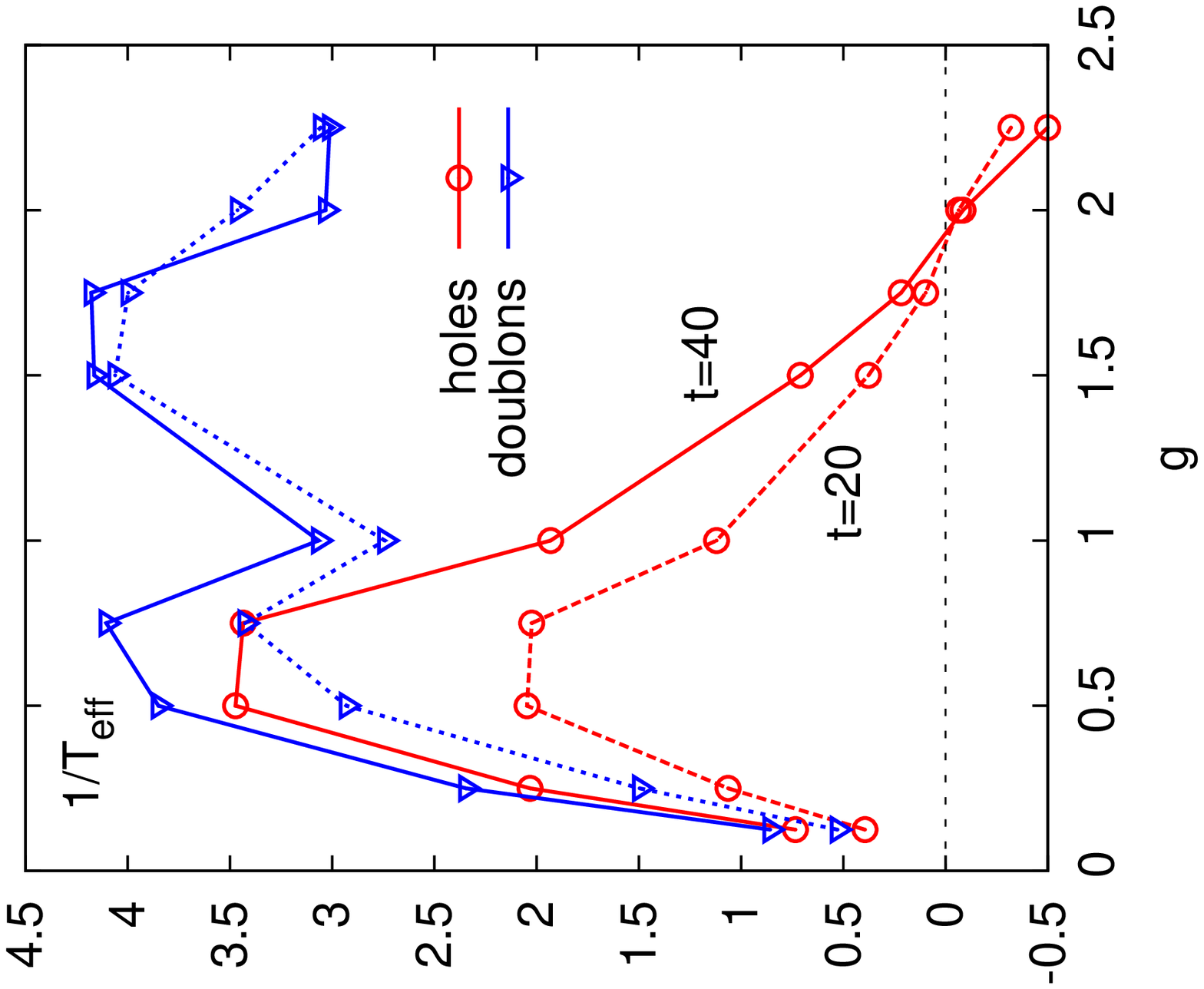}
\caption{Effective temperatures in systems with $U=15$ and photo-doping concentration $0.1$. Left panel: constant effective coupling strength $\lambda=g^2/\omega_0=0.25$. Middle panel: constant boson frequency $\omega_0=1$.  The upper curves show the hole temperatures and the lower curves the doublon temperatures. Right panel: inverse effective temperatures for doublons and holes as a function of $g$. The dashed (full) curves show the result at $t=20$ ($t=40$). 
}
\label{cooling_gw}
\end{center}
\end{figure}

We also analyzed the dependence on the parameters $g$ and $\omega_0$ and found that the cooling rate of the doublons and holes is essentially determined by $\lambda=g^2/\omega_0$. This is illustrated in the left panel of Fig.~\ref{cooling_gw}, which shows the effective doublon and hole temperatures for $U=15$, photo-doping concentration $0.1$, and different parameter choices corresponding to $\lambda=0.25$. The rate with which the effective temperature decreases is similar in the three cases, even though the boson frequencies and coupling strengths are very different, and the slowest boson oscillation period ($\approx 100$) is longer than the plotted time range.  

\begin{figure}[t]
\begin{center}
\includegraphics[angle=-90, width=0.49\columnwidth]{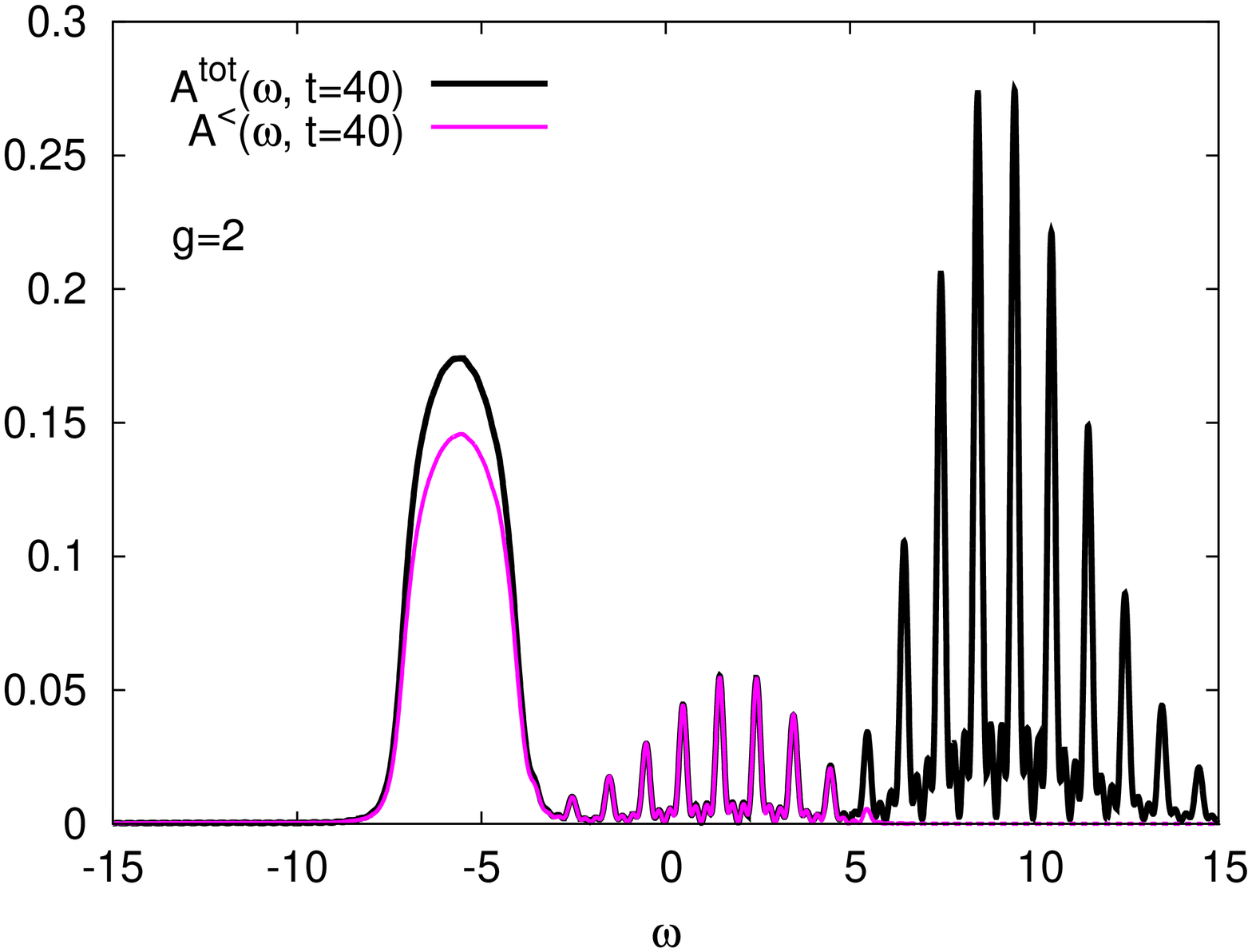}\hfill
\includegraphics[angle=-90, width=0.49\columnwidth]{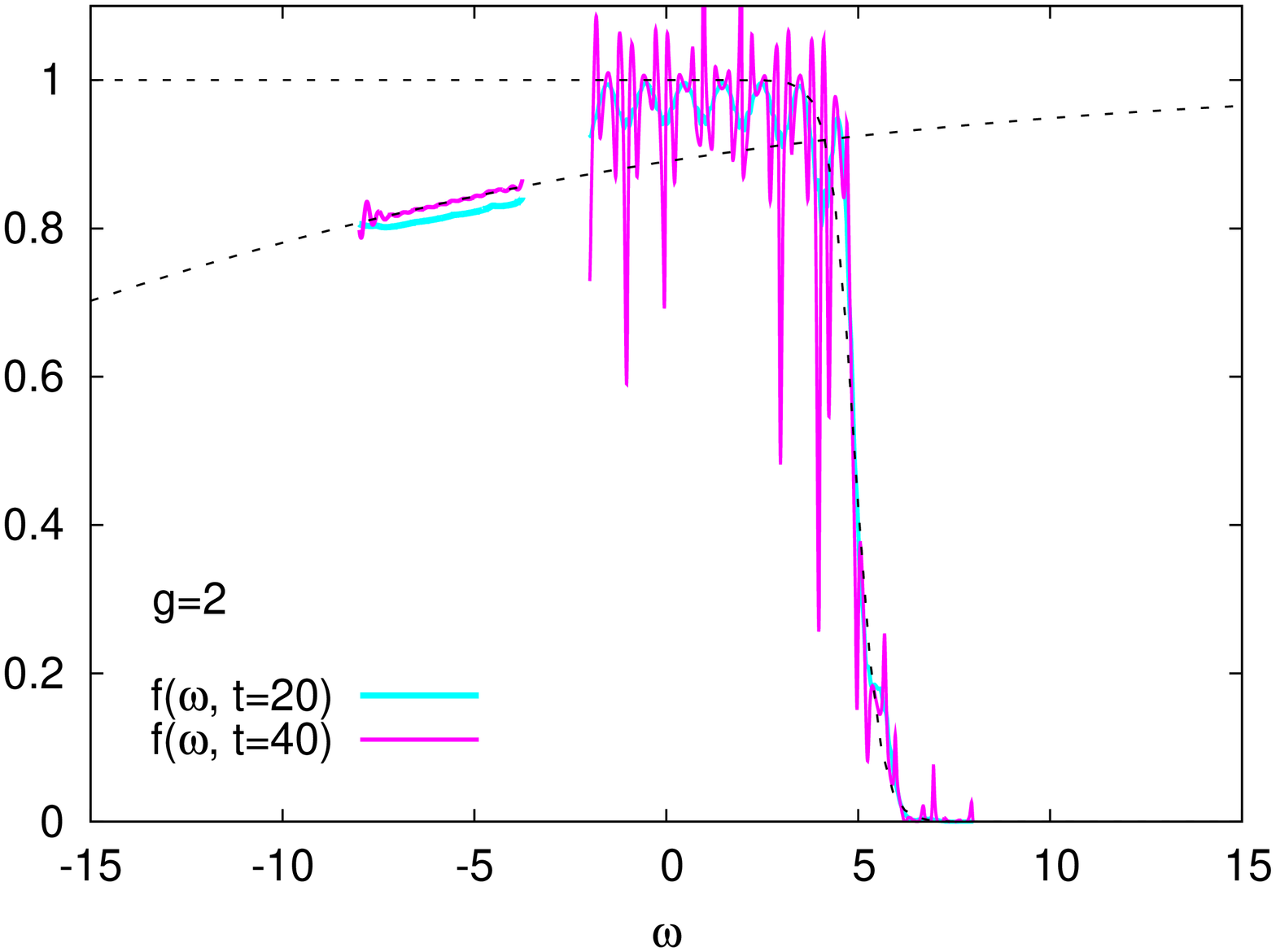}\\
\includegraphics[angle=-90, width=0.49\columnwidth]{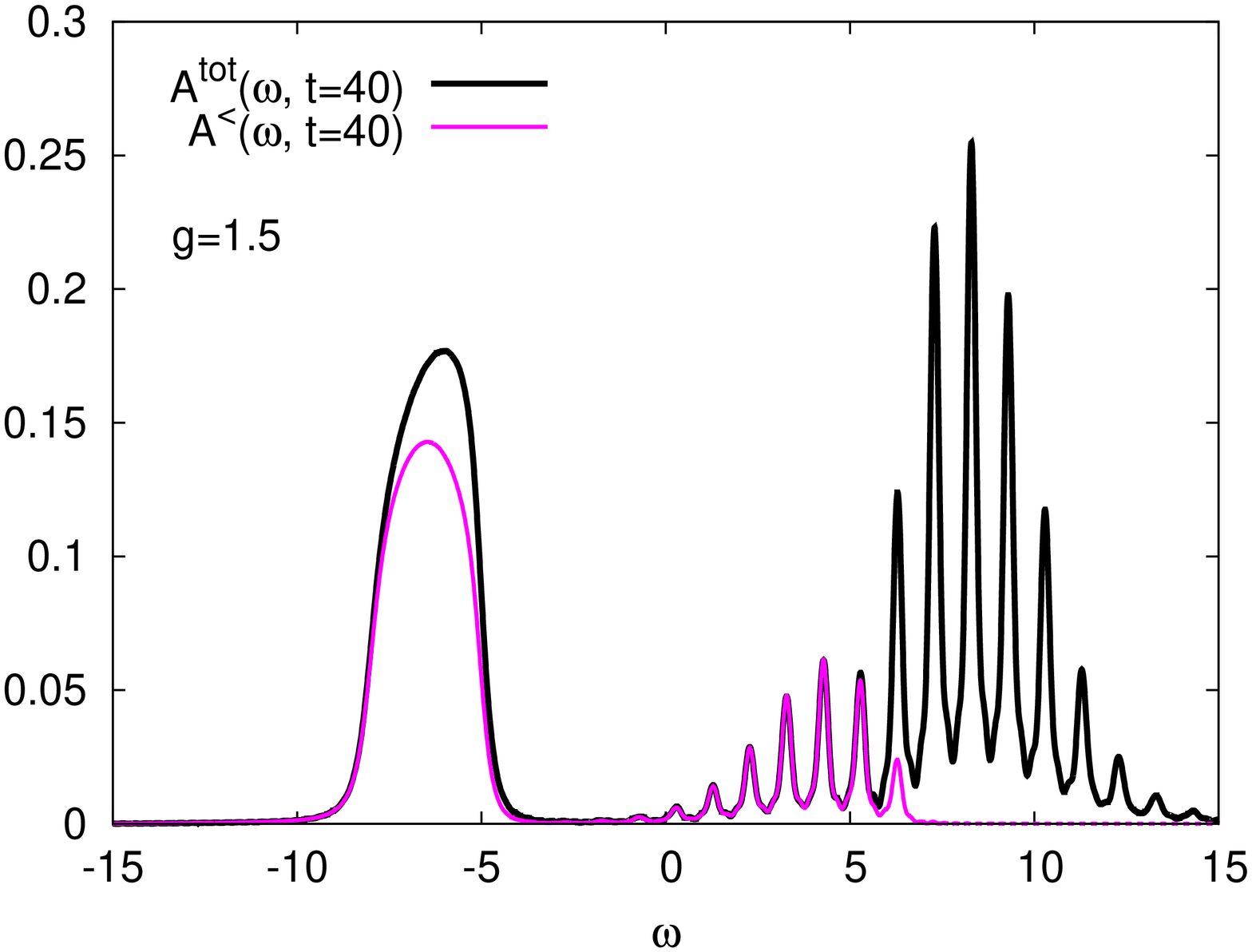}\hfill
\includegraphics[angle=-90, width=0.49\columnwidth]{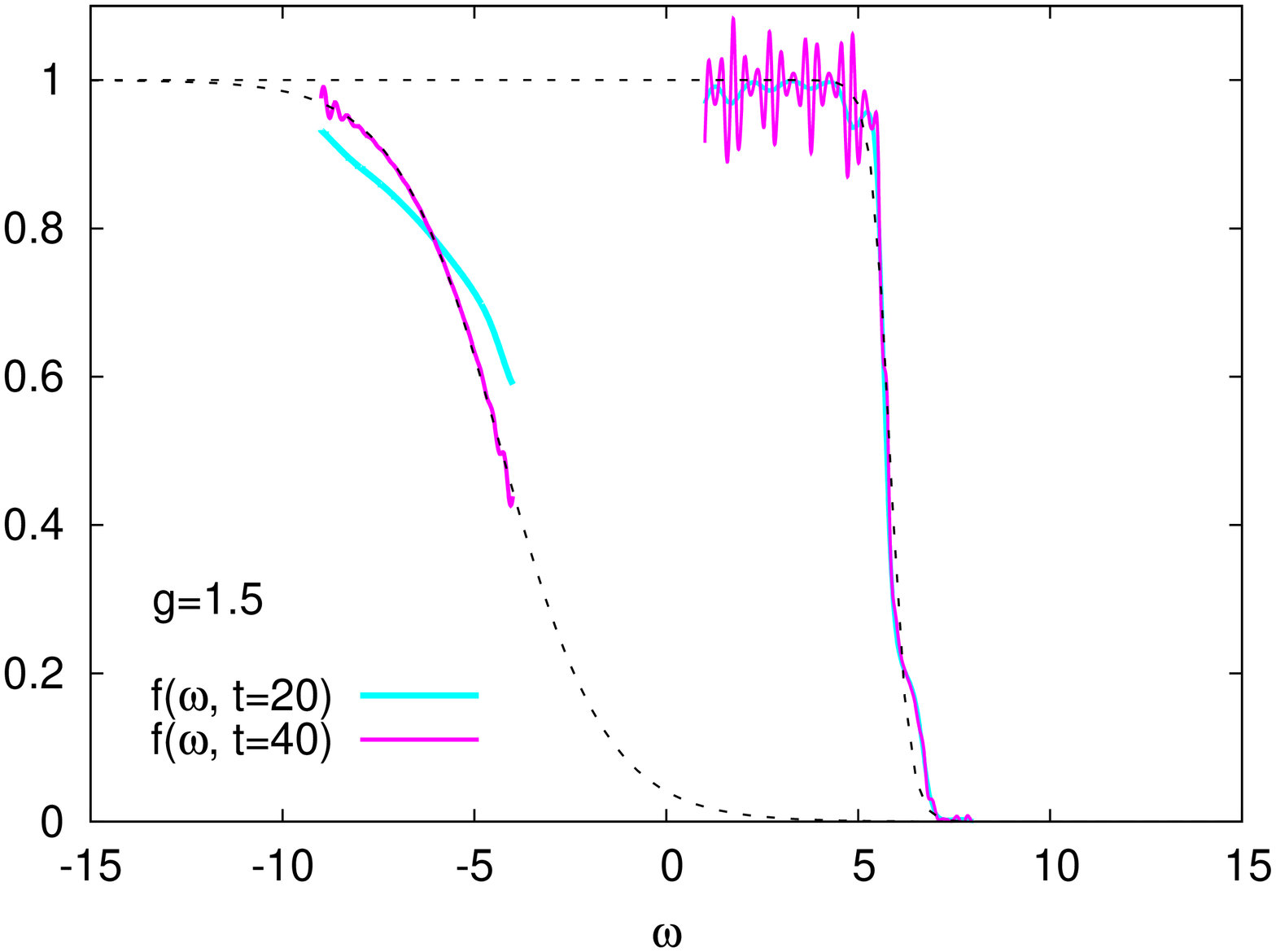}
\caption{Spectral functions and distribution functions in the polaronic regime ($U=15$, $\omega_0=1$, $g=2$ and $1.5$). In the right hand panels, we plot the distribution functions in the energy range corresponding to a sizable hole and doublon population. For the later time ($t=40$), we also show the Fermi function fits from which the effective doublon and hole temperatures have been extracted. For $g=2$, the holes exhibit an inverted population. The high frequency noise is due to the cutoff of the Fourier integral. 
}
\label{inverted}
\end{center}
\end{figure}

For fixed $\omega_0=1$, the cooling rate depends in a non-monotonous way on the coupling strength $g$ (see middle panel). If $g$ is small, the energy dissipation to the bosons becomes slow. As a result, the doublon temperature decreases slowly and the effective hole and doublon temperatures approach each other. In the intermediate coupling regime, the energy dissipation is most effective and both the doublon and hole temperatures decay quickly to a value close to the initial equilibrium temperature. In the polaronic regime, the doublons cool down very rapidly (although their effective temperature is no longer very well defined due to deviations of $f(\omega,t)$ from a Fermi distribution function, see Fig.~\ref{inverted}), while the transfer of kinetic energy from the holes to the doublons becomes increasingly inefficient, so that the holes remain hot. The inefficiency of the cooling via doublon-hole scattering is explained by the narrow width of the polaron sidebands, and the relatively large energy separation between them.  

The $g$-dependence of the inverse effective temperatures is shown in the right panel of Fig.~\ref{cooling_gw}. The most efficient cooling occurs around $g\approx 0.6$ (corresponding to $\lambda\approx 0.4$). For smaller $g$, the doublon and hole temperatures increase, while for larger couplings, only the hole temperature shows a marked increase. Interestingly, at $g\approx 1.9$, the holes reach an infinite temperature (flat) distribution, and for even stronger couplings, a negative temperature distribution (population inversion). From the comparison of the dashed and solid lines, which show the results for $t=20$ and $t=40$, it furthermore follows that the holes ``cool down" in this negative temperature state ($|T_\text{eff}|$ decreases), i.e., the inverted population becomes steeper with increasing time. 

Figure~\ref{inverted} shows the spectral functions and distribution functions for $g=1.5$ and $g=2$, corresponding to the polaronic regime with positive ($g=1.5$) and negative ($g=2$) effective hole temperature. By comparing the spectral functions for the two coupling strengths, one notices that in the $g=2$ case, the doublons populate in-gap states which fill the entire gap region. Due to these in-gap states, the doublon-hole recombination rate is substantially enhanced compared to $g=1.5$. 
A possible explanation for the inverted hole population is that the energy in a recombination processes is partially transferred to the hole subsystem, where it is stuck because holes do no longer efficiently thermalize with the doublons in the polaronic regime. The recombination process is still entropically favorable (although the ``entropy'' of the holons decreases with increasing negative temperatures) if a sufficient amount of energy is transferred to the bosons via the doublons. The rather long-lived transient state with inverted hole population thus results from doublon-hole recombinations, and the complicated energy exchange between holes, doublons and bosons in the polaronic regime with large $\omega_0$. On longer timescales, we expect that the system will thermalize in a state with identical, positive, temperatures for the doublons and holes.

\section{Summary and Outlook}
\label{summary}

We have discussed a DMFT based method for the simulation of nonequilibrium phenomena in the dynamic Hubbard model.\cite{Hirsch2001} This class of models has originally been introduced to capture the fact that atomic orbitals expand if occupied by two-electrons, which results in particle-hole asymmetric correlations. Dynamic Hubbard models have also recently appeared in the discussion of externally driven solids, where the orbital shapes are modulated periodically.\cite{Kaiser2014} We have considered here the model originally discussed by Hirsch, where the particle-hole asymmetric correlations are generated by a term $gdX$, which couples the doublon number operator to the position operator of an auxiliary boson.  

One can think of numerous nonequilibrium set-ups in which the particle-hole asymmetry leads to interesting effects, which are absent in a conventional Hubbard model or Holstein-Hubbard model description. While the external driving of the boson in a homogeneous Holstein-Hubbard model would result in a trivial time-evolution, the driving of the boson in the dynamic Hubbard model may produce  nontrivial effects, ranging from particle-hole excitations to dynamically induced boson-magnon couplings. If a temperature gradient is applied across a slab, the unequal doublon and hole mobilities should result in a nonzero particle current. 

In the present work we have focused our attention on photo-doped paramagnetic Mott insulators, where the asymmetric boson coupling leads to the rapid cooling of doublons and a hot distribution of holes, since the holes can dissipate their energy to the bosons only indirectly, via doublon-hole scattering. We have shown that a short time after a photo-doping excitation, the doublon and hole distributions in the upper and lower Hubbard band exhibit an approximately thermal shape, so that one can extract an effective doublon and hole temperature. We have analyzed the time evolution of these effective temperatures as a function of the model parameters. As in the Holstein-Hubbard model, the cooling rate seems to be determined by $\lambda=g^2/\omega_0$, and essentially independent of $U$. The fastest decrease of the effective hole temperature is observed for intermediate coupling strength, while in the polaronic regime, the energy transfer from the hot holes to the trapped doublons becomes slow. If the energy spacing $\omega_0$ between the polaronic states is large enough, the photo-doping of the Mott insulator may even lead to a long-lived transient state with an inverted hole population. 

A long-lived 
negative temperature state 
could lead to potentially interesting effects. In the weakly correlated metallic regime, it has been shown that a system with a dynamically generated population inversion behaves as if the Coulomb interaction had been switched from repulsive to attractive.\cite{Tsuji2011} While the present situation with strongly correlated photo-doped holes is quite different, one might 
nevertheless expect different phases to occur for $T>0$ and $T<0$ when $|T|$ becomes small. Whether or not the inverted hole population leads to symmetry breaking to unconventional ordered states is an interesting open issue. 

In the future, apart from studying driven systems and inhomogeneous set-ups, it will be interesting to extend the nonequilibrium DMFT simulations to other types of electron-boson couplings, such as the $gdX^2$ coupling of Ref.~\onlinecite{Kaiser2014}. The extension of the Lang-Firsov scheme to these variants of the dynamic Hubbard model is not straight-forward and may require additonal approximations. An alternative route is the development of combined strong/weak-coupling impurity solvers along the lines of Ref.~\onlinecite{Golez2015}.

\acknowledgements

We thank D. Golez and H. Strand for stimulating discussions. The calculations were run on the Beo04 cluster at the University of Fribourg. PW was supported by FP7 ERC Starting Grant No. 278023.

\end{document}